\documentclass{article}
\usepackage[utf8]{inputenc}
\usepackage[a4paper,left=3.5cm, right=3.5cm,top=3.5cm, bottom=3.5cm]{geometry}
\usepackage{amsmath}
\usepackage{amssymb}
\usepackage{graphicx}
\usepackage{esint}
\usepackage{mathtools, bm}
\usepackage{stmaryrd} 
\usepackage{slashed}
\usepackage{hyperref}
\hypersetup{
  colorlinks   = true,    
  urlcolor     = blue,    
  linkcolor    = blue,    
  citecolor    = black     
}

\usepackage{tikz}
\usetikzlibrary{positioning,arrows}
\usetikzlibrary{decorations.pathmorphing}
\usetikzlibrary{decorations.markings}
\tikzset{
particle/.style={thick,draw=black, postaction={decorate},
    decoration={markings,mark=at position .5 with {\arrow[black]{triangle 45}}}},
gluon/.style={decorate, draw=black,
    decoration={coil,aspect=0.8,segment length=4pt,amplitude=3pt}}
}

\newcommand{\explicitx}{x}
\newcommand{\Section}[1]{Section~\ref{#1}}
\newcommand{\Figure}[1]{Fig.~\ref{#1}}
\newcommand{\Equation}[1]{Eq.~(\ref{#1})}

\newcommand{\spa}[1]{\langle #1 \rangle}
\newcommand{\spb}[1]{[ #1 ]}    

\usepackage{color}
\usepackage[makeroom]{cancel}
\usepackage[normalem]{ulem}
\definecolor{pkcolor}{rgb}{0,0.1,0.7}

\newcommand\pkout{\marginpar{\color{pkcolor}$\clubsuit$}\bgroup\markoverwith{\color{pkcolor}{\rule[0.4ex]{2pt}{0.8pt}}}\ULon}

\definecolor{kkcolor}{rgb}{1,0,0}

\newcommand\kkout{\marginpar{\color{kkcolor}$\clubsuit$}\bgroup\markoverwith{\color{kkcolor}{\rule[04ex]{2pt}{0.8pt}}}\ULon}

\definecolor{avhcolor}{rgb}{1.0,0.3,0}

\newcommand\avhout{\marginpar{\color{avhcolor}$\clubsuit$}\bgroup\markoverwith{\color{avhcolor}{\rule[04ex]{2pt}{0.8pt}}}\ULon}

\title{%
\hspace{\fill}{\normalsize IFJPAN-IV-2020-6}\\[2ex]
All-plus helicity off-shell gauge invariant multigluon amplitudes at one loop}
\author{Etienne Blanco$^a$,
Andreas van Hameren$^a$,\\
Piotr Kotko$^b$,
Krzysztof Kutak$^a$
\\ \\
$^a${\it Institute of Nuclear Physics, Polish Academy of Sciences} \\
{\it  Radzikowskiego 152, 31-342 Krakow, Poland } \\ \\
$^b${\it AGH University Of Science and Technology, Physics Faculty,} \\ 
{\it Mickiewicza 30, 30-059 Krakow, Poland} \\ \\
}

\date{}

\begin{document}
\maketitle

\begin{abstract}
    We calculate one loop scattering amplitudes for arbitrary number of positive helicity on-shell gluons and one off-shell gluon treated within the quasi-multi Regge kinematics. The result is fully gauge invariant and possesses the correct on-shell  limit. Our method is based on embedding the off-shell process, together with contributions needed to retain gauge invariance, in a bigger fully on-shell process with auxiliary quark or gluon line. 
\end{abstract}

\section{Introduction}
\label{sec:Intro}

Despite the high energy limit of Quantum Chromodynamics (QCD) (see eg.  \cite{Kovchegov:2012mbw} for a review) has been studied for over forty years, the confrontation of various small-$x$ approaches and experimental data is still not fully conclusive (here $x\sim 1/\sqrt{s}$ is the longitudinal fraction of hadron momentum carried by a parton and $s$ is the center-of-mass energy). On one hand, the experimental data relevant to the small-$x$ regime can be often explained by the collinear factorization, supplemented however with parton showers or other type of resummations and multi-parton interactions. On the other hand, certain types of reactions, for example the Mueller-Navalet jet production  \cite{Mueller:1986ey} give strong hints towards the need of inclusion of the small-$x$ effects \cite{Ducloue:2013bva}. In addition, collisions of protons with heavy nuclei provide further hints, as observed for instance in \cite{vanHameren:2019ysa} for the forward dijet production case. 

In order to provide more solid statements regarding the need of  small-$x$ approaches, one needs higher order corrections for various components of small-$x$ calculations, in particular for high energy partonic amplitudes. As a matter of fact, in collinear factorization, any partonic amplitude can be at present calculated at NLO automatically using computer software%
. This is still to be achieved in the small-$x$ domain and our work is a step forward towards that goal.

The key result in the small-$x$ field is the Balitsky-Fadin-Kuraev-Lipatov (BFKL) equation \cite{Kuraev:1977fs, Balitsky:1978ic}, which describes evolution in energy (or $x$) of the gluon Green function in the high energy limit. It can also be converted to the energy evolution of so-called unintegrated parton distribution functions, that unlike collinear PDFs, explicitly depend on the parton transverse momentum $k_T$. Other key results in the small-$x$ QCD constitute the $k_T$-factorization (called also high energy factorization)  \cite{Catani:1990eg,Collins:1991ty}, as well as further developments that overcome the unitarity bound violation by the BFKL equation and lead to the nonlinear evolution of    Balitsky-Kovchegov (BK) equation \cite{Balitsky:1995ub,Kovchegov:1999yj}, B-JIMWLK equations \cite{Balitsky:1995ub,
JalilianMarian:1997jx,JalilianMarian:1997gr,JalilianMarian:1997dw,Kovner:2000pt,Kovner:1999bj,Weigert:2000gi,Iancu:2000hn,Ferreiro:2001qy} and Color Glass Condensate (CGC) effective theory (see e.g. \cite{Gelis:2010nm}). Some key higher order results include: the next-to-leading order (NLO) BFKL kernel \cite{Fadin:1998py,Ciafaloni:1998gs,Kotikov:2000pm}, the NLO BK kernel \cite{Balitsky:2008zza}, the B-JIMWLK equation at NLO \cite{Balitsky:2013fea,Kovner:2013ona}, the $\gamma^*\rightarrow \bar{q}q$ impact factor at NLO \cite{Bartels:2002uz,Balitsky:2012bs,Beuf:2016wdz,Boussarie:2016ogo} also with heavy quarks \cite{Chachamis:2013bwa}, partial inclusion of NLO for Higgs + jet \cite{Celiberto:2020tmb}, single inclusive jet production in CGC at NLO \cite{Chirilli:2011km}, and also the recent calculation of $\gamma^*\rightarrow \bar{q}q\gamma$ impact factor at NLO \cite{Roy:2019hwr}. In addition, there are NLO calculations in the context of the Lipatov's effective action \cite{Hentschinski:2011tz,Chachamis:2012gh,Chachamis:2012cc,Hentschinski:2014lma,Nefedov:2017qzc,Nefedov:2019mrg}.

The concept of $k_T$-factorization is based on analogy with the collinear factorization, but here both a hard part and a soft hadronic part depend on theparton transverse momenta, i.e.\ we have explicit higher powers $k_T/Q$ present in the hard matrix elements (here, $Q$ is the largest scale present in the process). Thus, instead of the leading twist, the accuracy is set by the leading power in $1/\sqrt{s}$. The momenta of partons defining the hard amplitude may now be off-shell, with vector or spinor indices projected onto components dominating in the high energy limit. 

\begin{figure}
    \centering
    \includegraphics[width=8.cm]{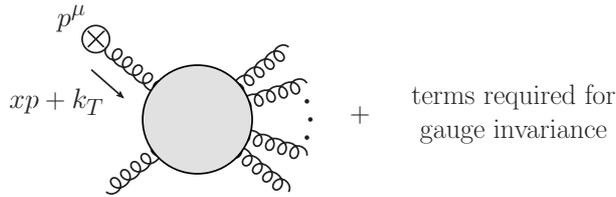}
    \caption{
    \small 
    In high energy factorization for forward jets (hybrid factorization \cite{Dumitru:2005gt,Deak:2009xt}) the multigluon amplitude has one incoming momentum off mass shell, with the off-shell propagator projected onto
    light-like momentum $p^\mu$ (typically the momentum of the hadron to which the gluon couples).
    The momentum of the off-shell leg has only one longitudinal component in the high energy kinematics. Such amplitude is in general not gauge invariant and additional terms are required to define it properly.
    }
    \label{fig:one-leg-offshell}
\end{figure}

In the present work we shall consider multigluon amplitudes with a single gluon being off mass shell. Such amplitudes are primarily used in the forward particle production (see eg. \cite{Deak:2009xt}) and have large phenomenological impact (see eg.  \cite{vanHameren:2013fla,vanHameren:2014ala,vanHameren:2014lna,vanHameren:2016ftb,Bury:2016cue,Bury:2017xwd,Kotko:2017oxg,Mantysaari:2019nnt,VanHaevermaet:2020rro,VanHaevermaet:2020rro} for various application in forward jet production processes at LHC). The momentum of the off-shell gluon has the form
\begin{equation}
    k^{\mu} = xp^{\mu} + k_T^{\mu} \, ,
    \label{eq:HEF_kinem}
\end{equation}
where $p^\mu$ is the light-like momentum typically associated with the colliding hadron, $x$ is the fraction of this momentum carried by the scattering parton, and $k_T^\mu$ is the transverse component satisfying $k_T\cdot p=0$. The off-shell gluon couples eikonally, i.e.\ its vector index is projected onto $p^\mu$ (the propagator is included in the amplitude), see Fig.~\ref{fig:one-leg-offshell}. The standard diagrams contributing to off-shell amplitude defined in that fashion are however not gauge invariant. The proper definition of such amplitudes can be done either within the Lipatov's high energy effective action \cite{Lipatov:1995pn,Antonov:2004hh} or by explicitly constructing additional contributions required by the gauge invariance, the  high energy kinematics and the proper soft and collinear behavior. The latter method is very useful in automated calculations at tree level and a few approaches exist: using the Ward identities \cite{vanHameren:2012uj}, embedding the off-shell process in a bigger on-shell one \cite{vanHameren:2012if} (see also \cite{Leonidov:1999nc} for earlier application to $2\rightarrow 2$ process), using matrix elements of straight infinite Wilson lines \cite{Kotko:2014aba}. In particular, the embedding method \cite{vanHameren:2012if} has proved to be very effective in numerical calculations and is implemented in a Monte Carlo generator \cite{vanHameren:2016kkz}. Also, it has been generalized to one-loop level with a proof of concept given in \cite{vanHameren:2017hxx}. 
The great advantage of this method is that it can be used to extract the high energy off-shell amplitudes from existing one loop on-shell results. We will review the method in detail in Section~\ref{sec:Method}.

In order to apply the embedding method at one-loop level, and in particular to validate the general concept of \cite{vanHameren:2017hxx}, it is reasonable to start with the simplest one-loop helicity amplitudes. In the on-shell case, these are the amplitudes with all helicities being the same, say 'plus' (we use the convention that all momenta are outgoing). Such amplitudes vanish at tree level, but are non-zero at loop level. Thus, in the present work, we shall calculate one-loop amplitudes with all-plus helicity gluons and one off-shell gluon, consistent with the gauge invariance and the high energy limit of QCD. Our result will be presented for arbitrary number of gluons $n$. In particular, we find that for $n=3$ our general result coincides with the existing result obtained from Lipatov's effective action \cite{Nefedov:2019mrg}.

As a basis for our calculation we shall use the existing one-loop results for $(-+\dots +)$ helicity on-shell amplitudes, where the first pair of particles is either gluon pair or quark-antiquark pair.  The particles with helicity $+-$ will provide an auxiliary quark or gluon line, with corresponding external spinors parametrized in a way that -- upon taking a proper limit -- will guarantee both the high energy kinematics \Equation{eq:HEF_kinem} and the eikonal coupling for the internal off-shell gluon attached to it. 

We shall focus on the so-called color ordered amplitudes  corresponding to planar diagrams and utilize the spinor helicity method (see \cite{Mangano:1990by} for a review). At tree level, the color decomposition of a full gluon amplitude into color ordered amplitudes is
\begin{equation}
    \mathcal{M}^{a_1,\dots,a_n}_{\lambda_1,\dots,\lambda_n}\left(k_1,\dots,k_n\right) =
    \sum_{\textsf{perm.}(2\cdots n)}\, \mathrm{Tr}\left(t^{a_1}t^{a_2}\dots t^{a_n}\right)\, \mathcal{A}\left(1^{(\lambda_1)},2^{(\lambda_2)},\dots,n^{(\lambda_n)}\right) \, ,
    \label{eq:ColorDecomposition}
\end{equation}
where $t^a$ are color generators, $k_i$ is momentum of $i$-th gluon with helicity projection $\lambda_i$ and the sum goes over all non-cyclic permutations of the arguments of the trace and the arguments of the color ordered amplitudes $\mathcal{A}$. At one-loop level, additional double trace terms are present. They can be however obtained as linear combinations of the leading trace contributions.

It is known that the on-shell $(\pm+\dots +)$ one-loop amplitudes have rather simple structure, given by a rational function of spinor products. Consider for instance the all-plus on-shell leading trace color ordered amplitude. It has a remarkably simple form for arbitrary number of gluons (conjectured by Z. Bern, G. Chalmers, L. J. Dixon and D. A. Kosower in \cite{bern1993new, bern1994one} and demonstrated by G. Mahlon in \cite{mahlon1994multigluon}) :
\begin{equation}
\mathcal{A}_n^{(1)} = g_s^n\sum_{1\leq i<j<k<l \leq n}
\frac{\spa{ij}\spb{jk}\spa{kl}\spb{li}}{\spa{12}\cdots\spa{n1}}
\label{eq:Allplus_intro}
\end{equation}
Above, the spinor products are defined as
\begin{equation}
  \spa{ij} = \overline{u}_-\left(k_i\right)u_+\left(k_j\right) \qquad 
  \spb{ij} = \overline{u}_+\left(k_i\right)u_-\left(k_j\right) 
  \, ,
    \label{eq:spinorprod}    
\end{equation}
where $u_{\pm}(k)$ are the spinors of helicity $\pm$ for an on-shell momentum $k$.
The above result is most easily understood within the unitarity methods (see eg. \cite{Bern_2011}), or -- more generally -- the   on-shell methods (see \cite{Arkani-Hamed_book_2016} for a comprehensive review). 
The off-shell gauge invariant amplitudes we calculate in the present work inherit the rational structure.

Our paper has the following structure. In the next section, we will describe the embedding method in more detail. Next, in Section~\ref{sec:Results}, we will present the main results for the amplitudes. In Section~\ref{sec:Verif} we recalulate the amplitudes using the embedding method with auxiliary gluon line as a verification of our results. In Section~\ref{sec:Limit} we will investigate the on-shell limit of the obtained off-shell amplitudes. Finally, in Section~\ref{sec:Summary} we shall summarize our work and discuss further perspectives.

\section{The method}
\label{sec:Method}

The method to obtain the off-shell amplitudes we are about to use has been developed in \cite{vanHameren:2012if}. Here we shall apply it to obtain one loop scattering amplitudes for arbitrary number of positive helicity on-shell gluons and one off-shell gluon with the high energy kinematics \Equation{eq:HEF_kinem} (called also the quasi-multi-Regge kinematics).

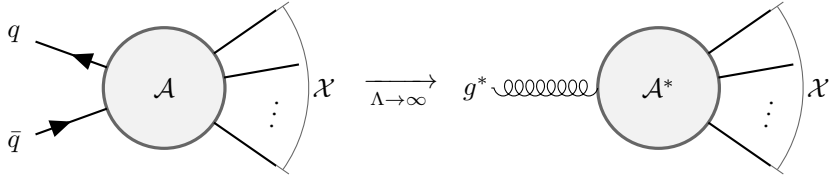
\begin{figure}
    \centering
\[\begin{tikzpicture}[baseline={(X.base)}]
 \node (X) at (0,0) {};
 \filldraw[color=black!60, fill=black!5, very thick](0,0) circle (.8);
 \draw[particle] (160:.8) -- (160:1.8);
 \node[overlay] at (160:2.1){$q$};
 \draw[particle] (-160:1.8) -- (-160:.8);
 \node[overlay] at (-160:2.1){$\bar{q}$};
 \node[overlay] at (0:0){$\mathcal{A}$};
 \foreach \X in {-35,10,35}
  {\draw[thick] (\X:.8) -- (\X:1.8);}
 \node [rotate=77.5,overlay] at (-12.5:1.5){$\cdots$};
 \draw[color=black!60,|-|] ([shift=(-35:1.9)]0,0) arc (-35:35:1.9);
 \node[overlay] at (0:2.1){$\mathcal{X}$};
\end{tikzpicture}
\hspace{.5cm}%
\xrightarrow[\Lambda\to \infty]{}%
\hspace{.6cm}
\begin{tikzpicture}[baseline={(X.base)}]
 \node (X) at (0,0) {};
\filldraw[color=black!60, fill=black!5, very thick](0,0) circle (.8);
 \foreach \X in {180}
  {\draw[gluon] (\X:.8) -- (\X:2.2);
   \node[overlay] at (\X+0:2.4){$g^*$};
   \node[overlay] at (\X+0:0){$\mathcal{A}^*$};
  }
 \foreach \X in {-35,10,35}
  {\draw[thick] (\X:.8) -- (\X:1.8);}
 \node [rotate=77.5,overlay] at (-12.5:1.5){$\cdots$};
 \draw[color=black!60,|-|] ([shift=(-35:1.9)]0,0) arc (-35:35:1.9);
 \node[overlay] at (0:2.1){$\mathcal{X}$};
\end{tikzpicture}
\]
    \caption{
    \small
    Gauge invariant off-shell amplitudes can be obtained by considering a process with an auxiliary quark-antiquark pair, with momenta parametrized in terms of a parameter $\Lambda$ in such a way, that upon taking the limit $\Lambda\rightarrow\infty$ the coupling to the quark line becomes eikonal and the momentum of the off-shell gluon has the high energy form (see \Equation{eq:HEF_kinem}).}
    \label{fig:HEFlimit}
\end{figure}

Let us briefly recall how the method works. The basic idea is to calculate the amplitude with the off-shell gluon using an on-shell amplitude with an auxiliary quark-antiquark pair, which follows specific kinematics. Ultimately, the auxiliary quark and antiquark spinors are decoupled ensuring gauge invariance of the off-shell amplitude. Schematically, the method can be summarized as (see also Fig.~\ref{fig:HEFlimit})
\begin{equation}
\lim_{\Lambda\to\infty}\left(\frac{\explicitx|k_T|}{g_s\Lambda}\mathcal{A}\left(\bar{q}(k_1)q(k_2)\mathcal{X}\right)\right)
=\mathcal{A}^*\left(g^*(k)\mathcal{X}\right)\, ,
\label{eq:lambdalimit}
\end{equation}
where $\mathcal{X}$ stands for other on-shell particles involved in the hard scattering process and $\Lambda$ is a real parameter parametrizing momenta of the auxiliary quark pair (see below).  The gauge invariant off-shell amplitude is denoted $\mathcal{A}^*$.
The momenta of the auxiliary quarks are taken to be the following:
\begin{equation}
\begin{split}
k_1^\mu &=  \Lambda p^\mu + \alpha q^\mu + \beta k_T^\mu \, ,\\
k_2^\mu &= k^\mu - k_1^\mu \, ,
\end{split}
\end{equation}
where
\begin{equation}
\alpha = \frac{-\beta^2k_T^2}{2\Lambda p \cdot q}
\quad ,\quad
\beta = \frac{1}{1+\sqrt{1-x/\Lambda}} 
\end{equation}
and $q^\mu$ is an arbitrary light-like momentum such that $q\cdot k_T=0$, $q\cdot p>0$.
Note, that $k_1^\mu$ and $k_2^\mu$ are light-like and they satisfy  
$k_1^\mu+k_2^\mu=k^\mu$, where the latter is the momentum of the off-shell gluon as defined in \Equation{eq:HEF_kinem}.
In the limit $\Lambda\to\infty$ the coupling of gluons to the quark line becomes eikonal, consistent with the high energy limit.
The factor $1/g_s$ in \Equation{eq:lambdalimit} is to correct the power of the coupling, and the factor $\explicitx|k_T|$ is for the correct matching to $k_T$-dependent PDFs in a cross section. In particular, the factor $|k_T|$ makes sure the amplitude is finite for $|k_T|\to0$.

 In practice, instead of using the above definitions of $k_1^\mu$ and $k_2^\mu$, we will use their expansion in $\Lambda$:
\begin{equation}
\begin{split}
k_1^\mu &= \Lambda p^\mu + \left(\frac{1}{2}+\frac{x}{8\Lambda}\right)k_T^\mu - \frac{k_T^2}{8\Lambda p \cdot q}q^\mu + \mathcal{O}(\Lambda^{-2}) \, ,
\\
k_2^\mu &= 
(x-\Lambda)p^\mu + \left(\frac{1}{2}-\frac{x}{8\Lambda}\right)k_T^\mu
+\frac{k_T^2}{8\Lambda p \cdot q}q^\mu + \mathcal{O}(\Lambda^{-2}) \, . \\
\end{split}
\end{equation}
In order to use the helicity method, we need to express $k_T^\mu$ in terms of spinors. It can be decomposed as follows
\begin{equation}
k_T^\mu = -\bar{\kappa}e^\mu - \bar{\kappa}^*e_*^\mu \, ,
\label{eq:kT1}
\end{equation}
with 
\begin{equation}
e^\mu=\frac{1}{2}\langle p|\gamma^\mu|q]
\quad,\quad
e_*^\mu=\frac{1}{2}\langle q|\gamma^\mu|p]
\label{eq:kT2}
\end{equation}
and
\begin{equation}
\bar{\kappa}=\frac{\kappa}{\spb{pq}}=
\frac{\langle q|\slashed{k}|p]}{2p \cdot q}
\quad,\quad
\bar{\kappa}^*=\frac{\kappa^*}{\spa{qp}}=
\frac{\langle p|\slashed{k}|q]}{2p \cdot q} \, .
\label{eq:kT3}
\end{equation}
Realize that $k_T^\mu$ is a four-vector with a negative square, and we have
\begin{equation}
    k_T^2 = -\kappa\kappa^*
    ~.
\end{equation}
The spinors of $k_1^\mu$ and $k_2^\mu$ can be decomposed into those of $p^\mu$ and $q^\mu$ following
\begin{align}
|1\rangle = \sqrt{\Lambda}\,|p\rangle 
          - \frac{\beta\bar{\kappa}^*}{\sqrt{\Lambda}}\,|q\rangle
\quad&,\quad
|1] = \sqrt{\Lambda}\,|p] 
    - \frac{\beta\bar{\kappa}}{\sqrt{\Lambda}}\,|q]\\
|2\rangle = \sqrt{\Lambda-x}\,|p\rangle 
          + \frac{\beta\bar{\kappa}^*}{\sqrt{\Lambda}}\,|q\rangle
\quad&,\quad
|2] = -\sqrt{\Lambda-x}\,|p] 
    - \frac{\beta\bar{\kappa}}{\sqrt{\Lambda}}\,|q]
    ~.
\end{align}
Notice that $\sqrt{(\Lambda-x)/\Lambda}\,\beta=1-\beta$. We see that the spinor products
\begin{equation}
\langle12\rangle = -\kappa^*
\quad,\quad
[12] = -\kappa
\end{equation}
are independent of $\Lambda$. Further, the spinors for auxiliary quarks behave for large $\Lambda$ as
\begin{equation}
|1\rangle\to\sqrt{\Lambda}\,|p\rangle \;\;,\;\;
|1]\to\sqrt{\Lambda}\,|p] \;\;,\;\;
|2\rangle\to\sqrt{\Lambda}\,|p\rangle \;\;,\;\;
|2]\to-\sqrt{\Lambda}\,|p] \;\;.
\end{equation}

In what follows, we shall call the above kinematics (together with taking the limit $\Lambda\to\infty$) the ``$\Lambda$ prescription''. Applying it to an amplitude with auxiliary partons gives the gauge invariant off-shell amplitude.

Alternatively, the ``embedding'' method described above can be used with an auxiliary gluon line, instead of the quark line. Indeed, the color decomposition for $(n-2)$-gluon amplitude with a quark-antiquark pair is given by
\begin{equation}
    \mathcal{M}^{ij\hspace{3.8ex}a_3,\dots,a_n}_{\lambda_1,\lambda_2,\lambda_3,\dots,\lambda_n}\left(k_1,\dots,k_n\right) =
    \sum_{\textsf{perm.}(3\cdots n)}\, \left(t^{a_3}\cdots t^{a_n}\right)_{ij}\, \mathcal{A}\left(q^{(\lambda_1)},\bar{q}^{(\lambda_2)},3^{(\lambda_3)},\dots,n^{(\lambda_n)}\right) \, ,
    \label{eq:ColorDecompositionQQ}
\end{equation}
and can be projected onto $(n-1)$-gluon amplitude by a contraction with $(t^{a_*})_{ji}$, where $a_*$  represents the color index of the off-shell gluon. Now, for an auxiliary gluon pair instead of a quark one, one simply needs to select only those permutations in \Equation{eq:ColorDecomposition} that retains the order of gluons $1$ and $2$ and substitute $t^{a_1}t^{a_2}\to t^{a_*}$.
At one loop, the color decompositions get more complicated and are given by Eq.~(1) in~\cite{bern1993one} and Eq.~(2.4-5) in~\cite{bern1995one} respectively. One can however easily see that the same procedure goes through to extract a single gluon color from a pair of colors.
In~\cite{vanHameren:2017hxx} it has been shown that at tree level, the partial amplitudes obtained using different pairs of auxiliary partons are identical. We will see here that the same holds at one loop for the all-plus amplitudes.

\section{All-plus off-shell gauge invariant amplitudes at NLO}
\label{sec:Results}

In this section we present our results for one loop amplitudes for one off-shell gluon and $n-1$ on-shell positive helicity gluons.  We begin with several low multiplicity examples, starting with the simplest cases: $n=3$ (the vertex), $n=4$ and $n=5$. Then, we will turn to a general result for arbitrary $n$. For each case, we first present the known amplitude with auxiliary quarks. Then, we apply the $\Lambda$ prescription to it and give the result for the off-shell amplitude.

\subsection{3-point vertex}
\label{sec:3ptResult}
We first consider the 3-point vertex with one off-shell gluon and two positive helicity on-shell gluons at one loop. Such vertex has been calculated for arbitrary helicity projection in \cite{Nefedov:2019mrg} from the Lipatov's effective action. 

In order to calculate it from the $\Lambda$ prescription, we need the 4-point amplitude for quark, anti-quark and two gluons. At one loop it has the following form \cite{bern1995one} :
\begin{equation}
\label{eq:ptqqAmp}
\mathcal{A}_4^{(1)}(1_{\bar{q}}^-,2_q^+,3^+,4^+) = 
-\frac{ig_s^4}{16\pi^2} \left[ 
\frac{1}{2}\left(1+\frac{1}{N_c^2}\right) +
\frac{1}{3}\left(1+\frac{n_s-n_f}{N_c}\right)\frac{s_{23}}{s_{12}}
\right]\frac{\spa{12}\spb{24}}{\spa{23}\spa{34}} \, ,
\end{equation}
where $n_f$ accounts for the number of Weyl fermions circulating in the loop, $n_s$ the number of complex scalars and
\begin{equation}
\forall i,j=1,\dots,n, \ s_{ij} = 2k_i\cdot k_j = \spa{ij}\spb{ji} \text{ (in this section, }n=4 \text{)}.
\end{equation}
Applying the $\Lambda$ prescription we get: 
\begin{equation}
\label{eq:3ptAmp}
\begin{split}
\mathcal{A}_3^{*(1)}(g^*,3^+,4^+) &= 
-\frac{ig_s^3}{24\pi^2}\left(1+\frac{n_s-n_f}{N_c}\right)
\explicitx|k_T|\frac{p\cdot k_3}{k_T^2}
\frac{\kappa^*\spb{p4}}{\spa{p3}\spa{34}}\\
&=  -\frac{ig_s^3}{24\pi^2}\left(1+\frac{n_s-n_f}{N_c}\right)
\frac{\explicitx|k_T|}{\kappa^2}p\cdot k_3
\frac{\spb{p3}\spb{p4}}{\spa{p3}\spa{p4}},
\end{split}
\end{equation}
where we used that $k_T^2=-\kappa\kappa^*$, $\kappa=\langle4|k\hspace{-1.2ex}/|p]/\langle4p\rangle$ and $k^\mu=-k_3^\mu-k_4^\mu$.
We checked that for $n_s=0$ the above result agrees with the one of \cite{nefedov2019one, Nefedov:2019mrg}, up to an overall constant and a factor $xE/|k_T|$, where $E$ is the energy component of $p^\mu$.
This difference is due to the fact that in the mentioned publications vertices rather than amplitudes are calculated (see also the comparisons at tree-level in~\cite{vanHameren:2012if}).

\subsection{4-point amplitude}
\label{sec:4ptResult}
The 5-leg amplitude with auxiliary quark pair is given by \cite{bern1995one} :
\begin{equation}
\label{eq:5ptqqAmp}
\begin{split}
\mathcal{A}_5^{(1)}(1_{\bar{q}}^-,2_q^+,3^+,4^+,5^+) = &
-\frac{ig_s^5}{32\pi^2}\left(1+\frac{1}{N_c^2}\right)\frac{\spa{12}\spb{23}\spa{31} + \spa{14}\spb{45}\spa{51}}{\spa{23}\spa{34}\spa{45}\spa{51}} \\
-\frac{ig_s^5}{48\pi^2}\left(1+\frac{n_s-n_f}{N_c}\right)&\bigg(
\frac{\spa{13}\spb{34}\spa{41}^2}{\spa{12}\spa{34}^2\spa{45}\spa{51}}
+\frac{\spa{14}\spa{24}\spb{45}\spa{51}}{\spa{12}\spa{23}\spa{34}\spa{45}^2} 
+\frac{\spb{23}\spb{25}}{\spb{12}\spa{34}\spa{45}}\bigg).
\end{split}
\end{equation}
Applying the $\Lambda$ prescription we find that the first term is of the order $\Lambda^{-1}$ and thus vanishes. Further calculation leads to the following result
\begin{equation}
\label{eq:4ptAmp}
\begin{split}
\mathcal{A}_4^{*(1)}(g^*,3^+,4^+,5^+) = &
-\frac{ig_s^4}{48\pi^2}\frac{\explicitx|k_T|\left(1+\frac{n_s-n_f}{N_c}\right)}
{\kappa^*\spa{p3}\spa{34}\spa{45}\spa{5p}} \\
& \times\bigg[\spa{p3}^2\spa{p4}^2\frac{\spb{34}}{\spa{34}}
+\spa{p4}^2\spa{p5}^2\frac{\spb{45}}{\spa{45}}
-\frac{\kappa^*}{\kappa}s_{p3}s_{p5}\bigg] \, .
\end{split}
\end{equation}

\subsection{5-point amplitude}
\label{sec:5ptResult}
The amplitude with the auxiliary quark pair is given by \cite{bern2005last} :
\begin{equation}
\label{eq:6ptqqAmp}
\begin{split}
\mathcal{A}_6^{(1)}(1_{\bar{q}}^-,2_q^+,3^+,4^+,5^+,6^+) = &
\frac{ig_s^6}{32\pi^2}\left(1+\frac{1}{N_c^2}\right)
\frac{\sum_{l=3}^5{\langle 1|\slashed{K}_{2\dots l}\slashed{k}_l}|1]}{\spa{23}\spa{34}\spa{45}\spa{56}\spa{61}} \\
+\frac{ig_s^6}{48\pi^2}\left(1+\frac{n_s-n_f}{N_c}\right)\Bigg[ &
\frac{\spa{14}\langle 1|(2+3)(3+4)|1\rangle}{\spa{12}\spa{34}^2\spa{45}\spa{56}\spa{61}}
+\frac{\spa{24}\spa{15}\langle 1|(4+5)(5+6)|1\rangle}{\spa{12}\spa{23}\spa{34}\spa{45}^2\spa{56}\spa{61}} \\
& - \frac{\spa{25}\langle 1|56|1\rangle}{\spa{12}\spa{23}\spa{34}\spa{45}\spa{56}^2}
+ \frac{\langle 1|3+4|2]^2}{\spa{34}^2\spa{56}\spa{61}\langle 5|3+4|2]} \\
& + \frac{\langle 2|4+5|6]\langle 1|4+5|6]^2}{\spa{12}\spa{23}\spa{45}^2\langle 3|4+5|6]s_{456}} \\
& - \frac{\spb{26}^2[2|(3+4)(4+5)(3+4)(4+5)|6]}{\spb{12}\spa{34}\spa{45}
\langle 5|3+4|2]\langle 3|4+5|6]s_{345}}\Bigg] \, ,
\end{split}
\end{equation}
where we defined
\begin{gather}
 \slashed{K}_{a\cdots b}=\sum_{i=a}^b\slashed{k}_i \,, \\
 s_{ijk}=\left(k_i+k_j+k_k\right)^2 \,,
 \end{gather}
and used
\begin{gather}
\langle a|(i+j)|b] = \spa{ai}\spb{ib} + \spa{aj}\spb{jb} \\
\langle a|(i+j)(k+l)|b\rangle
= \spa{ai}[i|(k+l)|b\rangle + \spa{aj}[j|(k+l)|b\rangle \,.
\end{gather}
Above $a,b,i,j,k,l=1,\dots n$, where  $n=6$ in the present section.

After we apply the $\Lambda$ prescription we find that the term with the factor $\left(1+\frac{1}{N_c^2}\right)$ vanishes leading to
\begin{equation}
\label{eq:5ptAmp}
\begin{split}
\mathcal{A}_5^{*(1)}(g^*,3^+,&4^+,5^+,6^+) = 
\left(1+\frac{n_s-n_f}{N_c}\right)\times\\
\times\frac{ig_s^5\explicitx|k_T|}{48\pi^2}\Bigg[&
\frac{\spa{p4}(\kappa^*[p|3+4|p\rangle+\spa{p3}\spb{34}\spa{4p})}{\kappa^*\spa{34}^2\spa{45}\spa{56}\spa{6p}}
+\frac{\spa{p4}\spa{p5}\langle p|(4+5)(5+6)|p\rangle}{\kappa^*\spa{p3}\spa{34}\spa{45}^2\spa{56}\spa{6p}} \\
& - \frac{\spa{p5}\langle p|56|p\rangle}{\kappa^*\spa{p3}\spa{34}\spa{45}\spa{56}^2}
+ \frac{\langle p|3+4|p]^2}{\spa{34}^2\spa{56}\spa{6p}\langle 5|3+4|p]} \\
& + \frac{\langle p|4+5|6]^3}{\kappa^*\spa{p3}\spa{45}^2\langle 3|4+5|6]s_{456}}
- \frac{\spb{p6}^2[p|(3+4)(4+5)(3+4)(4+5)|6]}{\kappa\spa{34}\spa{45}
\langle 5|3+4|p]\langle 3|4+5|6]s_{345}}\Bigg] \, .
\end{split}
\end{equation}

\subsection{$n$-point amplitude}
\label{sec:NptResult}

Finally, in the following section we shall derive the general expression for one-loop amplitude for one off-shell gluon and $n-1$ on-shell gluons with all helicities positive. To this end, we need 
the one loop amplitude for a quark-antiquark pair and $n-1$ positive helicity gluons. A suitable expression has been derived in \cite{bern2005last}. It reads
\begin{equation}
\label{eq:nptqqAmp}
\begin{split}
\mathcal{A}_{n+1}^{(1)}(1_{\bar{q}}^-,2_q^+,3^+, \cdots ,(n+1)^+) = &
\frac{ig_s^{n+1}}{32\pi^2}\left(1+\frac{1}{N_c^2}\right)
\frac{\sum_{l=3}^{n} \langle 1|\slashed{K}_{2 \cdots l}\slashed{k}_l|1\rangle}{\spa{23} \cdots \spa{(n+1)1}} \\
+& \frac{ig_s^{n+1}}{48\pi^2}\left(1+\frac{n_s-n_f}{N_c}\right)
\frac{S_1+S_2}{\spa{12}\spa{23} \cdots \spa{(n+1)1}} \, ,
\end{split}
\end{equation}
with 
\begin{equation}
\label{eq:nptAmpS}
\begin{split}
S_1 = & 
\sum_{j=3}^{n}\frac{\spa{2j}\spa{1(j+1)}\langle 1|\slashed{K}_{j,j+1}\slashed{K}_{(j+1) \cdots (n+1)}|1\rangle}{\spa{j(j+1)}} \, , \\
S_2 = &
\sum_{j=3}^{n-1}\sum_{l=j+1}^{n}
\frac{\langle 1|\slashed{K}_{j \cdots l}\slashed{K}_{(l+1) \cdots (n+1)}|1\rangle^2
\langle 2|\slashed{K}_{j \cdots l}\slashed{K}_{(l+1) \cdots (n+1)}|1\rangle}{\langle 1|\slashed{K}_{(l+1) \cdots (n+1)}\slashed{K}_{j \cdots l}|(j-1)\rangle
\langle 1|\slashed{K}_{(l+1) \cdots (n+1)}\slashed{K}_{j \cdots l}|j\rangle} \\
& \hspace{1.1cm} \times
\frac{\spa{(j-1)j}\spa{l(l+1)}
\langle 1|\slashed{K}_{2 \cdots (j-1)}[\mathcal{F}(j,l)]^2\slashed{K}_{(l+1) \cdots (n+1)}|1\rangle}
{\langle 1|\slashed{K}_{2 \cdots (j-1)}\slashed{K}_{j \cdots l}|l\rangle
\langle 1|\slashed{K}_{2 \cdots (j-1)}\slashed{K}_{j \cdots l}|(l+1)\rangle s_{j \cdots l}} \, , 
\end{split}
\end{equation}
where
\begin{equation}
\mathcal{F}(j,l) = 
\sum_{i=j}^{l-1}\sum_{m=i+1}^{l}\slashed{k}_i\slashed{k}_m \, .
\end{equation}
After applying the $\Lambda$ prescription we find that the term with the factor $\left(1+\frac{1}{N_c^2}\right)$ is of the order $\Lambda^{-1}$, whereas the other term is of order $1$ and is the one contributing to the off-shell amplitude. 
Eventually, we obtain the following expression for the off-shell amplitude:
\begin{equation}
\label{eq:nptAmp}
\mathcal{A}_n^{*(1)}(g^*,3^+, \cdots ,(n+1)^+) =
\frac{ig_s^n\explicitx|k_T|}{48\pi^2}\left(1+\frac{n_s-n_f}{N_c}\right)
\frac{U^*_1+U^*_2+U^*_3}{\kappa^*\spa{p3}\spa{34} \cdots \spa{np}} \, ,
\end{equation}
with
\begin{equation}
\label{eq:nptAmpU}
\begin{split}
U^*_1 = & 
\sum_{j=3}^{n}\frac{\spa{pj}\spa{p(j+1)}\langle p|\slashed{K}_{j,j+1}\slashed{K}_{(j+1) \cdots (n+1)}|p\rangle}{\spa{j(j+1)}} \, , \\
U^*_2 = &
\sum_{j=4}^{n-1}\sum_{l=j+1}^{n}
\frac{\langle p|\slashed{K}_{j \cdots l}\slashed{K}_{(l+1) \cdots (n+1)}|p\rangle^3}
{\langle p|\slashed{K}_{(l+1) \cdots (n+1)}\slashed{K}_{j \cdots l}|(j-1)\rangle
\langle p|\slashed{K}_{(l+1) \cdots (n+1)}\slashed{K}_{j \cdots l}|j\rangle} \\
& \hspace{1.1cm} \times
\frac{\spa{(j-1)j}\spa{l(l+1)}
\langle p|\slashed{K}'_{3 \cdots (j-1)}[\mathcal{F}(j,l)]^2\slashed{K}_{(l+1) \cdots (n+1)}|p\rangle}
{\langle p|\slashed{K}_{3 \cdots (j-1)}\slashed{K}_{j \cdots l}|l\rangle
\langle p|\slashed{K}_{3 \cdots (j-1)}\slashed{K}_{j \cdots l}|(l+1)\rangle s_{j \cdots l}} \, , \\
U^*_3 = &
\sum_{l=4}^{n}
\frac{\langle p|\slashed{K}_{3 \cdots l}\slashed{K}_{(l+1) \cdots (n+1)}|p\rangle^3}
{\langle p|\slashed{K}_{(l+1) \cdots (n+1)}\slashed{K}_{3 \cdots l}|p\rangle
\langle p|\slashed{K}_{(l+1) \cdots (n+1)}\slashed{K}_{3 \cdots l}|3\rangle} \\
& \hspace{1.1cm} \times
\frac{\spa{p3}\spa{l(l+1)}
[p|[\mathcal{F}(3,l)]^2\slashed{K}_{(l+1) \cdots (n+1)}|p\rangle}
{\kappa^*[p|\slashed{K}_{3 \cdots l}|l\rangle
[p|\slashed{K}_{3 \cdots l}|(l+1)\rangle s_{3 \cdots l}} \, .
\end{split}
\end{equation}

It can be readily checked that the above expression recovers the amplitudes calculated previously for $n=3,4,5$ in an independent way.

\section{Verification with auxiliary gluon pair}
\label{sec:Verif}

In the following section we shall verify the off-shell gauge invariant amplitudes we obtained in the previous section by applying the  $\Lambda$ prescription to the corresponding $n$-point amplitude with an auxiliary gluon pair, instead of the auxiliary quark pair. This will provide a nontrivial check of our calculations.

\subsection{3-point amplitude}
\label{sec:3ptVerif}
The 4-point one loop amplitude for one negative helicity gluon and three positive helicity gluons is given by \cite{bern1995one}
\begin{equation}
\label{eq:4ptggAmp}
\mathcal{A}_4^{(1)}(1^-,2^+,3^+,4^+) = \frac{ig_s^4}{48\pi^2}\left(1+\frac{n_s-n_f}{N_c}\right)
\frac{\spa{24}\spb{24}^3}{\spb{12}\spa{23}\spa{34}\spb{41}} \, .
\end{equation}
After applying the $\Lambda$ prescription we indeed find   that it leads to the same result as before:
\begin{equation}
\mathcal{A}_3^{*(1)}(g^*,3^+,4^+) 
= -\frac{ig_s^3\explicitx|k_T|}{24\pi^2}\left(1+\frac{n_s-n_f}{N_c}\right)
\frac{p\cdot k_3}{k_T^2}
\frac{\kappa^*\spb{p4}}{\spa{p3}\spa{34}} \, ,
\end{equation}
where we used $p\cdot k_3=-p\cdot k_4$ since $0=p\cdot k = p\cdot(-k_3-k_4)$.

\subsection{4-point amplitude}
\label{sec:4ptVerif}
The 5-point on-shell gluon amplitude we need is given by  \cite{bern1993one}
\begin{equation}
\label{eq:5ptggAmp}
\begin{split}
\mathcal{A}_5^{(1)}(1^-,2^+,3^+,4^+,5^+) = &
\frac{ig_s^5}{48\pi^2}\frac{\left(1+\frac{n_s-n_f}{N_c}\right)}{\spb{12}\spa{23}\spa{34}\spa{45}\spb{51}}
\times\bigg[(s_{23}+s_{34}+s_{45})\spb{25}^2 \\
- \spb{24}\spa{43}\spb{35}\spb{25}
-\frac{\spb{12}\spb{15}}{\spa{12}\spa{15}} & \left(
\spa{12}^2\spa{13}^2\frac{\spb{23}}{\spa{23}}+\spa{13}^2\spa{14}^2\frac{\spb{34}}{\spa{34}}+\spa{14}^2\spa{15}^2\frac{\spb{45}}{\spa{45}}\right)\bigg] \, .
\end{split}
\end{equation}
Applying the same procedure as before leads  to the following off-shell amplitude
\begin{equation}
\begin{split}
\mathcal{A}_4^{*(1)}(g^*,3^+,4^+,5^+) = &
\frac{ig_s^4}{48\pi^2}\frac{\explicitx|k_T|\left(1+\frac{n_s-n_f}{N_c}\right)}{\kappa^*\spa{p3}\spa{34}\spa{45}\spb{5p}}
\times\bigg[s_{p3}\spb{p5}^2 \\
-\frac{\kappa\spb{p5}}{\kappa^*\spa{p5}} & \left(
\spa{p3}^2\spa{p4}^2\frac{\spb{34}}{\spa{34}}+\spa{p4}^2\spa{p5}^2\frac{\spb{45}}{\spa{45}}\right)\bigg] \, ,
\end{split}
\end{equation}
which turns out to be equal to \Equation{eq:4ptAmp}.

\subsection{5-point amplitude}
\label{sec:5ptVerif}
In order to derive 5-point off-shell amplitude we use the following 6-point on-shell one loop amplitude \cite{bern2005last} :
\begin{equation}
\label{eq:6ptggAmp}
\begin{split}
\mathcal{A}_6^{(1)}(1^-,2^+,3^+,4^+,& 5^+,6^+) =
\frac{ig_s^6}{48\pi^2}\left(1+\frac{n_s-n_f}{N_c}\right)\times \\
\times\Bigg[ &
\frac{\langle 1|2+3|6]^3}{\spa{12}\spa{23}\spa{45}^2 s_{123}\langle 3|1+2|6]}
+\frac{\langle 1|3+4|2]^3}{\spa{34}^2\spa{56}\spa{61} s_{234}\langle 5|3+4|2]}\\ & +
\frac{\spb{26}^3}{\spb{12}\spb{61}s_{345}}\left(
\frac{\spb{23}\spb{34}}{\spa{45}\langle 5|3+4|2]}
-\frac{\spb{45}\spb{56}}{\spa{34}\langle 3|1+2|6]}
+\frac{\spb{35}}{\spa{34}\spa{45}}\right) \\
& - \frac{\spa{13}^3\spb{23}\spa{24}}{\spa{23}^2\spa{34}^2\spa{45}\spa{56}\spa{61}}
+\frac{\spa{15}^3\spa{46}\spb{56}}{\spa{12}\spa{23}\spa{34}\spa{45}^2\spa{56}^2} \\
& - \frac{\spa{14}^3\spa{35}\langle 1|2+3|4]}{\spa{12}\spa{23}\spa{34}^2\spa{45}^2\spa{56}\spa{61}} \Bigg] \, .
\end{split}
\end{equation}

Applying the $\Lambda$ prescription to the above on-shell result gives
\begin{equation}
\begin{split}
\mathcal{A}_5^{*(1)}(g^*,3^+,4^+,5^+,6^+) & = 
\frac{ig_s^5\explicitx|k_T|}{48\pi^2}\left(1+\frac{n_s-n_f}{N_c}\right)\times \\
\times\Bigg[ &
\frac{(\kappa^*\spb{p6}+\spa{p3}\spb{36})^3}{\kappa^*\spa{p3}\spa{45}^2 s_{k3}\langle 3|k|6]}
+\frac{\langle p|3+4|p]^3}{\spa{34}^2\spa{56}\spa{6p}(s_{p3}+s_{p4})\langle 5|3+4|p]}\\
&+\frac{\spb{p6}^2}{\kappa^*s_{345}}\left(
\frac{\spb{p3}\spb{34}}{\spa{45}\langle 5|3+4|p]}
-\frac{\spb{45}\spb{56}}{\spa{34}\langle 3|k|6]}
+\frac{\spb{35}}{\spa{34}\spa{45}}\right) \\
& - \frac{\spa{p3}\spb{p3}\spa{p4}}{\spa{34}^2\spa{45}\spa{56}\spa{6p}}
+\frac{\spa{p5}^3\spa{46}\spb{56}}{\kappa^*\spa{p3}\spa{34}\spa{45}^2\spa{56}^2} \\
& - \frac{\spa{p4}^3\spa{35}(\kappa^*\spb{p4}+\spa{p3}\spb{34})}{\kappa^*\spa{p3}\spa{34}^2\spa{45}^2\spa{56}\spa{6p}}\Bigg] \, .
\end{split}
\end{equation}
This amplitude turns out to be equal to the one obtained with auxiliary quark line,  \Equation{eq:6ptqqAmp}. The comparison is detailed in Appendix~\ref{APP:5ptCalc}.

\subsection{$n$-point amplitude}
\label{sec:NptVerif}
For the general case of $n$-point amplitude, the on-shell gluonic amplitude is taken from \cite{bern2005last}
\begin{equation}
\mathcal{A}_{n+1}^{(1)}(1^-,2^+,3^+, \cdots ,(n+1)^+) =
\frac{ig_s^{n+1}}{48\pi^2}\left(1+\frac{n_s-n_f}{N_c}\right)
\frac{T_1+T_2}{\spa{12}\spa{23} \cdots \spa{n1}} \, ,
\end{equation}
with
\begin{equation}
\label{eq:nptAmpT}
\begin{split}
T_1 = & 
\sum_{j=2}^{n}\frac{\spa{1j}\spa{1(j+1)}\langle 1|\slashed{K}_{j,j+1}\slashed{K}_{(j+1) \cdots (n+1)}|1\rangle}{\spa{j(j+1)}} \, , \\
T_2 = &
\sum_{j=3}^{n-1}\sum_{l=j+1}^{n}
\frac{\langle 1|\slashed{K}_{j \cdots l}\slashed{K}_{(l+1) \cdots (n+1)}|1\rangle^3}
{\langle 1|\slashed{K}_{(l+1) \cdots (n+1)}\slashed{K}_{j \cdots l}|(j-1)\rangle
\langle 1|\slashed{K}_{(l+1) \cdots (n+1)}\slashed{K}_{j \cdots l}|j\rangle} \\
& \hspace{1.3cm} \times
\frac{\spa{(j-1)j}\spa{l(l+1)}
\langle 1|\slashed{K}_{2 \cdots (j-1)}[\mathcal{F}(j,l)]^2\slashed{K}_{(l+1) \cdots (n+1)}|1\rangle}
{\langle 1|\slashed{K}_{2 \cdots (j-1)}\slashed{K}_{j \cdots l}|l\rangle
\langle 1|\slashed{K}_{2 \cdots (j-1)}\slashed{K}_{j \cdots l}|(l+1)\rangle s_{j \cdots l}} \, .
\end{split}
\end{equation}
Applying the $\Lambda$ prescription to $T_2$ gives the same result as for $S_2$ in \Equation{eq:nptAmpS}. It turns out that $T_1$ is equal to $S_1$ within the $\Lambda$ description once you realize that the first term in the sum over $j$ in $T_1$ is of the  order $\Lambda^{-1}$. In the end, applying the $\Lambda$ prescription to $\bar{q}^-q^+g^+ \cdots g^+$ or $g^-g^+g^+ \cdots g^+$ gives the same expression, given by \Equation{eq:nptAmp}.

\section{On-shell limit}
\label{sec:Limit}

Now, that we have obtained the expression for $\mathcal{A}_n^{*(1)}(g^*,3^+, \cdots ,(n+1)^+)$, we should verify that in the on-shell limit, i.e.\ when $|k_T|\to0$, we obtain the one loop on-shell amplitude with the first gluon having the momentum $xp^\mu$.
We expect that the limit consists of the sum of the amplitudes for which the, now on-shell, gluon has either negative or positive helicity.
For tree-level amplitudes, this can be understood as follows. Firstly, at the on-shell limit, the contributions to the amplitude that dominate have a propagator with denominator $k_T^2=-\kappa\kappa^*$, and have exactly the form of the first term in \Figure{fig:one-leg-offshell}.
More precisely, they have the form
\begin{equation}
 \sqrt{2}\,p^\mu\,\frac{\explicitx|k_T|}{\kappa\kappa^*}\,J_\mu
\end{equation}
where we use the planar Feynman rules as in Eq.~(10) of~\cite{vanHameren:2014iua}, where $J_\mu$ represents the off-shell current, and where we included the factor $\explicitx|k_T|$ from the $\Lambda$-prescription.
Using the current conservation $k\cdot J=0$, we can see that projecting on $p^\mu$ is equivalent to projecting on $-k_T^\mu/x$.
Secondly, using \Equation{eq:kT1} to \Equation{eq:kT3}, we see that
\begin{equation}
k_T^\mu =
-\frac{\kappa}{\sqrt{2}}\,\varepsilon_-^\mu(p,q)
-\frac{\kappa^*}{\sqrt{2}}\,\varepsilon_+^\mu(p,q) \, ,
\end{equation}
with polarization vectors
\begin{equation}
\varepsilon_-^\mu(p,q) = \frac{\langle p|\gamma^\mu|q]}{\sqrt{2}\,[pq]}
\quad,\quad
\varepsilon_+^\mu(p,q) = \frac{\langle q|\gamma^\mu|p]}{\sqrt{2}\,\langle qp\rangle}
~.
\end{equation}
Thus we find
\begin{equation}
\lim_{|k_T|\to0}{\mathcal{A}_n^{*(0)}(g^*\mathcal{X})} =
\frac{|k_T|}{\kappa^*}
\mathcal{A}_{n}^{(0)}(g^-\mathcal{X})
+\frac{|k_T|}{\kappa}
\mathcal{A}_{n}^{(0)}(g^+\mathcal{X})
~,
\label{eq:treeOnhellLim}
\end{equation}
where $|k_T|/\kappa^*=e^{\mathrm{i}\phi}$ for some angle $\phi$, and $|k_T|/\kappa$ its complex conjugate, and where $\mathcal{A}_{n}^{(0)}(g^\pm\mathcal{X})=\varepsilon_\pm\cdot J$.
In~\cite{vanHameren:2014iua} it is explained how such a coherent sum of amplitudes becomes an incoherent sum of squared amplitudes in a cross section.

When taking the on-shell limits in expressions consisting of spinor products and invariants involving the momentum~$p^\mu$, the final step is to interpret this momentum as the momentum of the now on-shell gluon, divided by $x$. Since the tree amplitudes are homogeneous in $p^\mu$ of degree~$1$, this results in the overall factor $1/x$ equivalent to the one coming from changing the projector $p^\mu\to-k_T^\mu/x$ above. The off-shell one-loop all-plus amplitudes can easily be checked to be homogeneous in $p^\mu$ of degree~$1$ too, and the same factor $1/x$ will show up to cancel the factor $\explicitx$ from the $\Lambda$-prescription.

We now verify that the same limit appears for the one-loop $n$-point all-plus amplitudes we obtained in \Section{sec:NptResult}.
One can notice that $U^*_1\xrightarrow[|k_T|\to0]{}T^1$ and $U^*_2\xrightarrow[|k_T|\to0]{}T^2$, which implies
\begin{equation}
\begin{split}
\lim_{|k_T|\to0}{\mathcal{A}_n^{*(1)}(g^*,3^+, \cdots ,(n+1)^+)} = &
\frac{|k_T|}{\kappa^*}\mathcal{A}_{n}^{(1)}(xp^-,3^+, \cdots ,(n+1)^+)\\
&+\frac{ig_s^n\explicitx}{48\pi^2}\frac{\lim_{k_T\to0}{\left(U^*_3|k_T|/\kappa^*\right)}}{\spa{p3}\spa{34} \cdots \spa{(n+1)p}} \, .
\label{eq:limAnU3}
\end{split}
\end{equation}
So we already have the contribution from the amplitude with negative helicity gluon (the first term in the expression above). We now need to show that the second term is actually the contribution from the amplitude with a positive helicity gluon, i.e.
\begin{equation}
\mathcal{A}_n^{(1)}(1^+, \cdots ,n^+) = \frac{ig^n}{48\pi^2}\left(1+\frac{n_s-n_f}{N_c}\right)
\sum_{1\leq i<j<k<l \leq n}
\frac{\spa{ij}\spb{jk}\spa{kl}\spb{li}}{\spa{12}\cdots\spa{n1}}.
\label{eq:nptAllplus}
\end{equation}
To this end, we have to manipulate on the expression $U^*_3$. One can show that
\begin{equation}
\begin{split}
\label{eq:limU3}
U^*_3 \xrightarrow[|k_T|\to 0]{} \frac{\kappa^*}{\kappa\spb{p(n+1)}}
[p|[\mathcal{F}(3,n)]^2|(n+1)]
= & \frac{\kappa^*}{\kappa}\sum_{3\leq i<j<k<l\leq (n+1)}\spa{ij}\spb{jk}\spa{kl}\spb{li} \\
& + \frac{\kappa^*}{\kappa}\sum_{3\leq j<k<l\leq (n+1)}\spa{pj}\spb{jk}\spa{kl}\spb{lp} \, .
\end{split}
\end{equation}
Inserting this into \Equation{eq:limAnU3} leads to
\begin{equation}
\begin{split}
\lim_{k_T\to0}{\mathcal{A}_n^{*(1)}(g^*,3^+, \cdots ,(n+1)^+)} = &
\frac{|k_T|}{\kappa^*}\mathcal{A}_{n}^{(1)}(xp^-,3^+, \cdots ,(n+1)^+)  \\
& + \frac{|k_T|}{\kappa}\mathcal{A}_{n}^{(1)}(xp^+,3^+, \cdots ,(n+1)^+) \, .
\end{split}
\end{equation}
More details on the above rather non-trivial calculation are given in Appendix~\ref{APP:Collinear}.
This is exactly what we expect from the on-shell limit of an off-shell amplitude, based on the limit for tree-level amplitudes (\Equation{eq:treeOnhellLim}): a superposition of on-shell amplitudes, where the off-shell gluon is replaced by a gluon with a positive and negative helicity.

\section{Summary}
\label{sec:Summary}
In this paper we have calculated expressions for amplitudes in high energy factorization with one off-shell gluon and any number of plus-helicity gluons at one loop level.
We also obtained expressions for specific cases: 3, 4 and 5 point amplitudes. 
To obtain these results we used the embedding method developed in \cite{vanHameren:2012if,vanHameren:2017hxx}. The method relies on identifying pair of on-shell partons as auxiliary lines which can be decoupled in high energy limit, leaving gauge invariant off-shell amplitude with proper high energy kinematics. 
We find agreement with the existing calculation for the 3-point vertex with a Reggeized gluon in~\cite{Nefedov:2019mrg}.
Furthermore we explicitly demonstrated that we obtain the correct on-shell limit for all calculated amplitudes. Thus, we conclude that the embedding method works
at the one-loop level, at least for amplitudes with same helicities.

Our future plans involve calculation of other QCD amplitudes and, in particular, addressing also the real corrections. The ultimate goal is to automatize the NLO calculations in $k_T$-factorization as well as the small-$x$ improved TMD factorization (ITMD) \cite{Kotko:2015ura,Altinoluk:2019fui}.

\section*{Acknowledgments}
E. Blanco, A. van Hameren and K. Kutak acknowledge partial support by NCN grant No. DEC-2017/27/B/ST2/01985. P. Kotko is supported by the NCN grant No. DEC-2018/31/D/ST2/02731.

\appendix
\section{5-point amplitude -- detailed calculation}
\label{APP:5ptCalc}

In order to compare the off-shell gauge invariant 5-point amplitude obtained from the auxiliary quark line $\bar{q}^-q^+g^+g^+g^+g^+$ to the one obtained from the auxiliary gluon line $g^-g^+g^+g^+g^+g^+$, we will rewrite both expressions. Let's first rewrite the first term of the amplitude with auxiliary quarks (before applying the $\Lambda$ prescription, see \Equation{eq:6ptqqAmp})
\begin{equation}
\begin{split}
\frac{\spa{14}\langle 1|(2+3)(3+4)|1\rangle}{\spa{12}\spa{34}^2\spa{45}\spa{56}\spa{61}}
&=\frac{\spa{14}(\spa{12}\spb{23}\spa{31}+\langle 1|2+3|4]\spa{41})}{\spa{12}\spa{34}^2\spa{45}\spa{56}\spa{61}} \\
&=-\frac{\spa{14}\spb{23}\spa{13}}{\spa{34}^2\spa{45}\spa{56}\spa{61}}
+\frac{\spa{23}\spa{45}\spa{14}^2\langle 1|5+6|4]}{\spa{12}\spa{23}\spa{34}^2\spa{45}^2\spa{56}\spa{61}} \\
&=-\frac{\spa{14}\spb{23}\spa{13}}{\spa{34}^2\spa{45}\spa{56}\spa{61}}
+\frac{\spa{24}\spa{35}\spa{14}^2\langle 1|5+6|4]}{\spa{12}\spa{23}\spa{34}^2\spa{45}^2\spa{56}\spa{61}} \\
&-\frac{\spa{25}\spa{14}^2\langle 1|5+6|4]}{\spa{12}\spa{23}\spa{34}\spa{45}^2\spa{56}\spa{61}} \, . \\
\end{split}
\end{equation}
Above, we have used the momentum conservation to write $\langle 1|2+3|4]=-\langle 1|5+6|4]$ and the Schouten identity: $\spa{23}\spa{45}=\spa{24}\spa{35}+\spa{25}\spa{43}$. 
It leads to
\begin{equation}
\begin{split}
&\mathcal{A}_5^{(1)}(g^*,3^+,4^+,5^+,6^+) \\
&= \frac{ig^5\explicitx|k_T|}{48\pi^2}\Bigg[
 -\frac{\spa{p4}\spb{p3}\spa{p3}}{\spa{34}^2\spa{45}\spa{56}\spa{6p}}
 +\frac{\spa{35}\spa{p4}^3\langle p|5+6|4]}{\kappa^*\spa{p3}\spa{34}^2\spa{45}^2\spa{56}\spa{6p}} \\
&\hspace{13.5ex}
-\frac{\spa{p5}\spa{p4}^2\langle p|5+6|4]}{\kappa^*\spa{p3}\spa{34}\spa{45}^2\spa{56}\spa{6p}} 
+ \frac{\langle p|4+5|6]^3}{\kappa^*\spa{p3}\spa{45}^2\langle 3|4+5|6]s_{k3}} \\
&\hspace{13.5ex}
 +\frac{\spa{p4}\spa{p5}(\spa{p4}[4|5+6|p\rangle+\spa{p5}\spb{56}\spa{6p})}{\kappa^*\spa{p3}\spa{34}\spa{45}^2  \spa{56}\spa{6p}}
- \frac{\spa{p5}\langle p|56|p\rangle}{\kappa^*\spa{p3}\spa{34}\spa{45}\spa{56}^2} \\
&\hspace{13.5ex}
 + \frac{\langle p|3+4|p]^2}{\spa{34}^2\spa{56}\spa{6p}\langle 5|3+4|p]}
- \frac{\spb{p6}^2[p|(3+4)(4+5)(3+4)(4+5)|6]}{\kappa\spa{34}\spa{45}
   \langle 5|3+4|p]\langle 3|4+5|6]s_{k6}}\Bigg]\\
&= \frac{ig^5\explicitx|k_T|}{48\pi^2}\Bigg[
 -\frac{\spa{p4}\spb{p3}\spa{p3}}{\spa{34}^2\spa{45}\spa{56}\spa{6p}}
 +\frac{\spa{35}\spa{p4}^3\langle p|5+6|4]}{\kappa^*\spa{p3}\spa{34}^2\spa{45}^2\spa{56}\spa{6p}} \\
&\hspace{13.5ex}
+\frac{\langle p|4+5|6]^3}{\kappa^*\spa{p3}\spa{45}^2\langle 3|4+5|6]s_{k3}}
 +\frac{\spa{p4}\spa{p5}^2\spb{56}\spa{6p}}{\kappa^*\spa{p3}\spa{34}\spa{45}^2\spa{56}\spa{6p}} \\
&\hspace{13.5ex}
-\frac{\spa{p5}^2\spb{56}\spa{6p}}{\kappa^*\spa{p3}\spa{34}\spa{45}\spa{56}^2}
 + \frac{\langle p|3+4|p]^2}{\spa{34}^2\spa{56}\spa{6p}\langle 5|3+4|p]} \\
&\hspace{13.5ex}
- \frac{\spb{p6}^2[p|(3+4)(4+5)(3+4)(4+5)|6]}{\kappa\spa{34}\spa{45}
\langle 5|3+4|p]\langle 3|4+5|6]s_{k6}}\Bigg] \, .
\end{split}
\end{equation}
Terms 4 and 5 can be combined using the Schouten identity
\begin{equation}
\frac{\spa{p5}^2\spb{56}}{\kappa^*\spa{p3}\spa{34}\spa{45}^2\spa{56}^2}
(\spa{p4}\spa{56}-\spa{6p}\spa{45})
=\frac{\spa{p5}^3\spa{46}\spb{56}}{\kappa^*\spa{p3}\spa{34}\spa{45}^2\spa{56}^2} \, .
\end{equation}
Thus, finally, the amplitude reads
\begin{equation}\label{eq.Rggg_q}
\begin{split}
&\mathcal{A}_5^{(1)}(g^*,3^+,4^+,5^+,6^+) \\
&= \frac{ig^5\explicitx|k_T|}{48\pi^2}\Bigg[
\frac{\langle p|4+5|6]^3}{\kappa^*\spa{p3}\spa{45}^2\langle 3|4+5|6]s_{k3}}
+\frac{\langle p|3+4|p]^2}{\spa{34}^2\spa{56}\spa{6p}\langle 5|3+4|p]}\\
&\hspace{13.5ex}
-\frac{\spb{p6}^2[p|(3+4)(4+5)(3+4)(4+5)|6]}{\kappa\spa{34}\spa{45}
\langle 5|3+4|p]\langle 3|4+5|6]s_{k6}}
-\frac{\spa{p4}\spb{p3}\spa{p3}}{\spa{34}^2\spa{45}\spa{56}\spa{6p}} \\
&\hspace{13.5ex}
+\frac{\spa{p5}^3\spa{46}\spb{56}}{\kappa^*\spa{p3}\spa{34}\spa{45}^2\spa{56}^2}
+\frac{\spa{35}\spa{p4}^3\langle p|5+6|4]}{\kappa^*\spa{p3}\spa{34}^2\spa{45}^2\spa{56}\spa{6p}}\Bigg] \, .
\end{split}
\end{equation}

Let us now rewrite the expression for the amplitude \Equation{eq:6ptggAmp}. In the second term we use
\begin{equation}
s_{234}  = s_{23}+s_{24}+s_{34} \xrightarrow[\Lambda \text{ prescr.}]{}
\Lambda(s_{p3}+s_{p4}) + \mathcal{O}(1) = \Lambda(\spa{p3}\spb{3p}+\spa{p4}\spb{4p})
=\Lambda\langle p|3+4|p] \, .
\end{equation}
In the first term we use
\begin{equation}
\langle 1|2+3|6] = -\langle 1|4+5|6] \xrightarrow[\Lambda \text{ prescr.}]{}
-\Lambda\langle p|4+5|6] + \mathcal{O}(1)
\end{equation}
For the factorized term in the second line, we can use the momentum conservation
\begin{equation}
s_{345} = s_{k6} \, .
\end{equation}
For the last term, before applying $\Lambda$ prescription, we use :
\begin{equation}
\langle 1|2+3|4] = -\langle 1|5+6|4] \xrightarrow[\Lambda \text{ prescr.}]{}
-\Lambda\langle p|5+6|4] + \mathcal{O}(1)
\end{equation}

In the end, we have
\begin{equation}\label{eq.Rggg_g}
\begin{split}
&\mathcal{A}_5^{(1)}(g^*,3^+,4^+,5^+,6^+) \\
& = \frac{ig^5\explicitx|k_T|}{48\pi^2}\Bigg[
-\frac{\langle p|4+5|6]^3}{\kappa^*\spa{p3}\spa{45}^2 s_{k3}\langle 3|k|6]}
+\frac{\langle p|3+4|p]^2}{\spa{34}^2\spa{56}\spa{6p}\langle 5|3+4|p]}\\
&\hspace{13.5ex}
+\frac{\spb{p6}^2}{\kappa^*s_{k6}}\left(
\frac{\spb{p3}\spb{34}}{\spa{45}\langle 5|3+4|p]}
-\frac{\spb{45}\spb{56}}{\spa{34}\langle 3|k|6]}
+\frac{\spb{35}}{\spa{34}\spa{45}}\right) \\
&\hspace{13.5ex}
- \frac{\spa{p3}\spb{p3}\spa{p4}}{\spa{34}^2\spa{45}\spa{56}\spa{6p}}
+\frac{\spa{p5}^3\spa{46}\spb{56}}{\kappa^*\spa{p3}\spa{34}\spa{45}^2\spa{56}^2} \\
&\hspace{13.5ex}
+ \frac{\spa{p4}^3\spa{35}\langle p|5+6|4]}{\kappa^*\spa{p3}\spa{34}^2\spa{45}^2\spa{56}\spa{6p}} \Bigg] \, .
\end{split}
\end{equation}
Let us now compare \Equation{eq.Rggg_q} and \Equation{eq.Rggg_g}.
It is clear that the terms 2, 4, 5 and 6 are the same. The first terms are also equal upon applying $\langle 3|4+5|6]=-\langle 3|k|6]$. Let us now work on the third term of \Equation{eq.Rggg_q}:
\begin{equation}
\begin{split}
\frac{[p|(3+4)(4+5)(3+4)(4+5)|6]}{\spa{34}\spa{45}
\langle 5|3+4|p]\langle 3|4+5|6]}
=\frac{[5|(3+4)(4+5)|6]}{\spa{34}\spa{45}\langle 3|4+5|6]}
+\frac{[p|343(4+5)|6]}{\spa{34}\spa{45}\langle 5|3+4|p]\langle 3|4+5|6]}\\
=\frac{\spb{53}\langle 3|(4+5)|6]}{\spa{34}\spa{45}\langle 3|4+5|6]}
+\frac{\spb{54}\spa{45}\spb{56}}{\spa{34}\spa{45}\langle 3|4+5|6]}
+\frac{\spb{p3}\spa{34}\spb{43}\langle 3|(4+5)|6]}{\spa{34}\spa{45}
\langle 5|3+4|p]\langle 3|4+5|6]}\\
=-\frac{\spb{35}}{\spa{34}\spa{45}}
+\frac{\spb{45}\spb{56}}{\spa{34}\langle 3|k|6]}
-\frac{\spb{p3}\spb{34}}{\spa{45}
\langle 5|3+4|p]} \, .
\end{split}
\end{equation}
If we put back the factor $-\frac{\spb{p6}^2}{\kappa^*s_{k6}}$ (not writen in the calculation for simplicity), we recognize the second line of \Equation{eq.Rggg_g}. Thus, both approaches give the same result.

\section{On-shell limit calculation}
\label{APP:Collinear}
In this appendix we detail the calculation that leads to \Equation{eq:limU3} which implies the correct on-shell limit for the $n$-point off-shell amplitude we presented in \Equation{eq:nptAmpT}.

In order to rewrite the expression for $U^*_3$ so that the on-shell limit can be utilized, let us come back to the expression for $T_2$, see \Equation{eq:nptAmpT} before applying the $\Lambda$ prescription. We focus on the first term in the sum over $j$  (i.e.\ for $j=3$), since it is the term that leads to $U^*_3$ when applying the $\Lambda$ prescription. Let us call this term $T_3$:
\begin{equation}
\begin{split}
T_3 = &
\sum_{l=4}^{n}
\frac{\langle 1|\slashed{K}_{3 \cdots l}\slashed{K}_{(l+1) \cdots (n+1)}|1\rangle^3}
{\langle 1|\slashed{K}_{(l+1) \cdots (n+1)}\slashed{K}_{3 \cdots l}|2\rangle
\langle 1|\slashed{K}_{(l+1) \cdots (n+1)}\slashed{K}_{3 \cdots l}|3\rangle} \\
& \hspace{1.1cm} \times
\frac{\spa{23}\spa{l(l+1)}\spa{12}
[2|[\mathcal{F}(3,l)]^2\slashed{K}_{(l+1) \cdots (n+1)}|1\rangle}
{\spa{12}[2|\slashed{K}_{3 \cdots l}|l\rangle
\spa{12}[2|\slashed{K}_{3 \cdots l}|(l+1)\rangle s_{3 \cdots l}} \, .
\end{split}
\end{equation}
We have
\begin{equation}
\langle 1|\slashed{K}_{3\dots l}\slashed{K}_{(l+1)\dots (n+1)}|1\rangle =
-\langle 1|\slashed{K}_{3\dots l}^2|1\rangle-\langle 1|\slashed{K}_{3\dots l}|2]\spa{21}
\xrightarrow[\Lambda \text{ prescr.}]{}\Lambda\kappa^*\sum_{i=3}^l s_{pi} \, .
\end{equation}
Similar, we have
\begin{equation}
\langle 1|\slashed{K}_{(l+1)\dots (n+1)}\slashed{K}_{3\dots l}|2\rangle =
-\spa{12}[2|\slashed{K}_{3\dots l}|2\rangle
\xrightarrow[\Lambda \text{ prescr.}]{}-\Lambda\kappa^*\sum_{i=3}^l s_{pi} \, ,
\end{equation}
\begin{equation}
\begin{split}
\langle 1|\slashed{K}_{(l+1)\dots (n+1)}\slashed{K}_{3\dots l}|3\rangle = &
-\langle 1|\slashed{K}_{3\dots l}^2|3\rangle-\spa{12}[2|\slashed{K}_{3\dots l}|3\rangle \\
= & \spa{13}s_{3\dots l}
-\langle 1|\slashed{K}_{3\dots l}^2|3\rangle-\spa{12}[2|\slashed{K}_{3\dots l}|3\rangle \, ,
\end{split}
\end{equation}
which implies
\begin{equation}
\langle 1|\slashed{K}_{(l+1)\dots (n+1)}\slashed{K}_{3\dots l}|3\rangle
\xrightarrow[\Lambda \text{ prescr.}]{}-\sqrt{\Lambda}\left(
\kappa^*[p|\slashed{K}_{3\dots l}|3\rangle
+\spa{p3}s_{3\dots l}\right)
\xrightarrow[k_T\to 0]{}-\sqrt{\Lambda}\spa{p3}s_{3\dots l} \, .
\end{equation}
We may also notice that, for $l=n$, we have
\begin{equation}
[2|\slashed{K}_{3\dots l}|(l+1)\rangle=[2|\slashed{K}_{3\dots n}|(n+1)\rangle
=-\spb{21}\spa{1(n+1)}\xrightarrow[\Lambda \text{ prescr.}]{}\kappa\spa{p(n+1)} \, .
\end{equation}
This is the only term in the sum over $l$ that has $\kappa$ in the denominator and that is the only non vanishing term when $k_T$ tends to 0. 

Putting all this together leads to
\begin{equation}
\begin{split}
T_3 \xrightarrow[\Lambda \text{ prescr.}]{} U^*_3 = &
\kappa^*\sum_{l=4}^{n-1}
\frac{\left(\sum_{i=3}^l s_{pi}\right)^2\spa{l(l+1)}
[p|[\mathcal{F}(3,l)]^2\slashed{K}_{(l+1) \cdots (n+1)}|p\rangle}
{[p|\slashed{K}_{3 \cdots l}|l\rangle
[p|\slashed{K}_{3 \cdots l}|(l+1)\rangle s_{3 \cdots l}^2} \\
& + \kappa^*\frac{\left(\sum_{i=3}^{n} s_{pi}\right)^2\spa{n(n+1)}
[p|[\mathcal{F}(3,n)]^2|(n+1)]\spa{(n+1)p}}
{\spb{p(n+1)}\spa{(n+1)n}\kappa\spa{p(n+1)}s_{3 \cdots n}^2} \\
& \xrightarrow[k_T\to 0]{} \frac{\kappa^*}{\kappa}
\frac{\left(\sum_{i=3}^{n} s_{pi}\right)^2[p|[\mathcal{F}(3,n)]^2|(n+1)]}
{\spb{p(n+1)}s_{3 \cdots n}^2} \, .
\end{split}
\end{equation}
Notice that
\begin{equation}
s_{3 \cdots n} = s_{p(n+1)} = \spa{p(n+1)}\spb{(n+1)p} = -\sum_{i=3}^{n}\spa{pi}\spb{ip}
= -\sum_{i=3}^{n}s_{pi} \, .
\end{equation}

Back to $U^*_3$, we have
\begin{equation}
\lim_{k_T\to0}{\frac{|k_T|}{\kappa^*}U^*_3} = 
\frac{|k_T|}{\kappa\spb{p(n+1)}}[p|[\mathcal{F}(3,n)]^2|(n+1)].
\end{equation}
This demonstrates the first relation in \Equation{eq:limU3}. We now have to prove the second one i.e.\ we need to show that the obtained expression corresponds to the numerator of the amplitude for $n-1$ gluons with positive helicity (up to some factor). Actually, we should first rewrite this numerator
\begin{equation}
\begin{split}
\label{eq:nptAmpggNum}
\sum_{1\leq i<j<k<l \leq n}\spa{ij}\spb{jk}\spa{kl}\spb{li}
= & \frac{1}{\spb{1n}}\sum_{1\leq i<j<k<l \leq n}
\spa{ij}\spb{jk}\spa{kl}\spb{li}\spb{1n} \\
= & \frac{1}{\spb{1n}}\sum_{1\leq i<j<k<l \leq n}
(\spb{1l}\spa{lk}\spb{kj}\spa{ji}\spb{in}
+\spb{1i}\spa{ij}\spb{jk}\spa{kl}\spb{ln}) \\
= & \frac{1}{\spb{1n}}\left(\sum_{2\leq i<j<k<l < n}
+\sum_{1\leq l<k<j<i \leq n}\right)
\spb{1i}\spa{ij}\spb{jk}\spa{kl}\spb{ln}.
\end{split}
\end{equation}

Now we can work on $U^*_3$. Let's first express $\mathcal{F}$ in terms of a sum. For a direct comparison, we should also use the expression of $U^*_3$ with the following change in the momenta label : 
$p\to 1$, $\forall i \in 3,\dots,n+1, \ i\to i-1$ (then momentum conservation expresses the same way i.e.\ $\sum_{i=1}^n k_i = 0$).
\begin{equation}
\begin{split}
\label{eq:U3limsum}
\lim_{k_T\to0}{\frac{|k_T|}{\kappa^*}U^*_3} & = 
\frac{|k_T|}{\kappa\spb{1n}}[1|[\mathcal{F}(2,n-1)]^2|n] \\
& = \frac{|k_T|}{\kappa\spb{1n}}
\sum_{\substack{2\leq i<j<n \\ 2\leq k<l<n}}
\spb{1i}\spa{ij}\spb{jk}\spa{kl}\spb{ln}
\end{split}
\end{equation} 
We have then for both \Equation{eq:U3limsum} and \Equation{eq:nptAmpggNum} a sum over the same expression. We can then, to shorten the demonstration, forget about the summed term (i.e.\ $\spb{1i}\spa{ij}\spb{jk}\spa{kl}\spb{ln}$) and work directly on the sums to show that they are the same in this context. \\
On one side we have
\begin{equation}
\begin{split}
\sum_{\substack{2\leq i<j<n \\ 2\leq k<l<n}} = &
\sum_{2\leq i<j<k<l<n}+\sum_{2\leq i\leq k<j\leq l<n}+\sum_{2\leq i\leq k<l<j<n} \\
&+\sum_{2\leq k<i<j\leq l<n}+\sum_{2\leq k<i\leq l<j<n}+\sum_{2\leq k<l<i<j<n}
\end{split}
\end{equation} 
on the other hand, we have
\begin{equation}
\resizebox{0.9\hsize}{!}{$
\begin{split}
\sum_{2\leq i<j<k<l<n}+\sum_{1\leq l<k<j<i\leq n} = &
\sum_{2\leq i<j<k<l<n}-\left(\sum_{2\leq k<l<j<i\leq n}
+\sum_{2\leq k<j\leq l<i\leq n}+\sum_{2\leq k<j<i\leq l<n}\right) \\
= & \sum_{2\leq i<j<k<l<n}+\left(\sum_{2\leq i\leq k<l<j<n}
+\sum_{2\leq k<i\leq l<j<n}+\sum_{2\leq k<l<i<j<n}\right) \\
& \hspace{2cm} +\left(\sum_{2\leq i\leq k<j\leq l<n}+\sum_{2\leq k<i<j\leq l<n}
+\sum_{2\leq k<j<i\leq l<n}\right)\\
& \hspace{2cm}-\sum_{2\leq k<j<i\leq l<n} \\
&= \sum_{\substack{2\leq i_1<j_1<n \\ 2\leq i_2<j_2<n}}
\end{split}$}
\end{equation}
The first equality is obtained by momentum conservation on the index $l$ (in the second term only) and the second one also by momentum conservation, on the index $i$ this time (for terms 2 and 3). This finally proves the second relation in \Equation{eq:limU3} which leads then to the expected on-shell limit for the amplitude in \Equation{eq:nptAmp}.


\bibliographystyle{JHEP}
\bibliography{references}{}

\providecommand{\href}[2]{#2}\begingroup\raggedright\begin{thebibliography}{10}

\bibitem{Kovchegov:2012mbw}
Y.~V. Kovchegov and E.~Levin, \emph{{Quantum chromodynamics at high energy}},
  vol.~33.
\newblock Cambridge University Press, 2012.

\bibitem{Mueller:1986ey}
A.~H. Mueller and H.~Navelet, \emph{{An Inclusive Minijet Cross-Section and the
  Bare Pomeron in QCD}},
  \href{http://dx.doi.org/10.1016/0550-3213(87)90705-X}{\emph{Nucl. Phys.}
  {\bfseries B282} (1987) 727--744}.

\bibitem{Ducloue:2013bva}
B.~Ducloué, L.~Szymanowski and S.~Wallon, \emph{{Evidence for high-energy
  resummation effects in Mueller-Navelet jets at the LHC}},
  \href{http://dx.doi.org/10.1103/PhysRevLett.112.082003}{\emph{Phys. Rev.
  Lett.} {\bfseries 112} (2014) 082003},
  [\href{https://arxiv.org/abs/1309.3229}{{\ttfamily 1309.3229}}].

\bibitem{vanHameren:2019ysa}
A.~van Hameren, P.~Kotko, K.~Kutak and S.~Sapeta, \emph{{Broadening and
  saturation effects in dijet azimuthal correlations in p-p and p-Pb collisions
  at $\mathbf{\sqrt{s}} = $ 5.02 TeV}},
  \href{http://dx.doi.org/10.1016/j.physletb.2019.06.055}{\emph{Phys. Lett. B}
  {\bfseries 795} (2019) 511--515},
  [\href{https://arxiv.org/abs/1903.01361}{{\ttfamily 1903.01361}}].

\bibitem{Kuraev:1977fs}
E.~A. Kuraev, L.~N. Lipatov and V.~S. Fadin, \emph{{The Pomeranchuk Singularity
  in Nonabelian Gauge Theories}}, {\emph{Sov. Phys. JETP} {\bfseries 45} (1977)
  199--204}.

\bibitem{Balitsky:1978ic}
I.~I. Balitsky and L.~N. Lipatov, \emph{{The Pomeranchuk Singularity in Quantum
  Chromodynamics}}, {\emph{Sov. J. Nucl. Phys.} {\bfseries 28} (1978)
  822--829}.

\bibitem{Catani:1990eg}
S.~Catani, M.~Ciafaloni and F.~Hautmann, \emph{{High-energy factorization and
  small x heavy flavor production}},
  \href{http://dx.doi.org/10.1016/0550-3213(91)90055-3}{\emph{Nucl. Phys.}
  {\bfseries B366} (1991) 135--188}.

\bibitem{Collins:1991ty}
J.~C. Collins and R.~K. Ellis, \emph{{Heavy quark production in very
  high-energy hadron collisions}},
  \href{http://dx.doi.org/10.1016/0550-3213(91)90288-9}{\emph{Nucl. Phys.}
  {\bfseries B360} (1991) 3--30}.

\bibitem{Balitsky:1995ub}
I.~Balitsky, \emph{{Operator expansion for high-energy scattering}},
  \href{http://dx.doi.org/10.1016/0550-3213(95)00638-9}{\emph{Nucl. Phys.}
  {\bfseries B463} (1996) 99--160},
  [\href{https://arxiv.org/abs/hep-ph/9509348}{{\ttfamily hep-ph/9509348}}].

\bibitem{Kovchegov:1999yj}
Y.~V. Kovchegov, \emph{{Small x F(2) structure function of a nucleus including
  multiple pomeron exchanges}},
  \href{http://dx.doi.org/10.1103/PhysRevD.60.034008}{\emph{Phys. Rev.}
  {\bfseries D60} (1999) 034008},
  [\href{https://arxiv.org/abs/hep-ph/9901281}{{\ttfamily hep-ph/9901281}}].

\bibitem{JalilianMarian:1997jx}
J.~Jalilian-Marian, A.~Kovner, A.~Leonidov and H.~Weigert, \emph{{The BFKL
  equation from the Wilson renormalization group}},
  \href{http://dx.doi.org/10.1016/S0550-3213(97)00440-9}{\emph{Nucl. Phys.}
  {\bfseries B504} (1997) 415--431},
  [\href{https://arxiv.org/abs/hep-ph/9701284}{{\ttfamily hep-ph/9701284}}].

\bibitem{JalilianMarian:1997gr}
J.~Jalilian-Marian, A.~Kovner, A.~Leonidov and H.~Weigert, \emph{{The Wilson
  renormalization group for low x physics: Towards the high density regime}},
  \href{http://dx.doi.org/10.1103/PhysRevD.59.014014}{\emph{Phys. Rev.}
  {\bfseries D59} (1998) 014014},
  [\href{https://arxiv.org/abs/hep-ph/9706377}{{\ttfamily hep-ph/9706377}}].

\bibitem{JalilianMarian:1997dw}
J.~Jalilian-Marian, A.~Kovner and H.~Weigert, \emph{{The Wilson renormalization
  group for low x physics: Gluon evolution at finite parton density}},
  \href{http://dx.doi.org/10.1103/PhysRevD.59.014015}{\emph{Phys. Rev.}
  {\bfseries D59} (1998) 014015},
  [\href{https://arxiv.org/abs/hep-ph/9709432}{{\ttfamily hep-ph/9709432}}].

\bibitem{Kovner:2000pt}
A.~Kovner, J.~G. Milhano and H.~Weigert, \emph{{Relating different approaches
  to nonlinear QCD evolution at finite gluon density}},
  \href{http://dx.doi.org/10.1103/PhysRevD.62.114005}{\emph{Phys. Rev.}
  {\bfseries D62} (2000) 114005},
  [\href{https://arxiv.org/abs/hep-ph/0004014}{{\ttfamily hep-ph/0004014}}].

\bibitem{Kovner:1999bj}
A.~Kovner and J.~G. Milhano, \emph{{Vector potential versus color charge
  density in low x evolution}},
  \href{http://dx.doi.org/10.1103/PhysRevD.61.014012}{\emph{Phys. Rev.}
  {\bfseries D61} (2000) 014012},
  [\href{https://arxiv.org/abs/hep-ph/9904420}{{\ttfamily hep-ph/9904420}}].

\bibitem{Weigert:2000gi}
H.~Weigert, \emph{{Unitarity at small Bjorken x}},
  \href{http://dx.doi.org/10.1016/S0375-9474(01)01668-2}{\emph{Nucl. Phys.}
  {\bfseries A703} (2002) 823--860},
  [\href{https://arxiv.org/abs/hep-ph/0004044}{{\ttfamily hep-ph/0004044}}].

\bibitem{Iancu:2000hn}
E.~Iancu, A.~Leonidov and L.~D. McLerran, \emph{{Nonlinear gluon evolution in
  the color glass condensate. 1.}},
  \href{http://dx.doi.org/10.1016/S0375-9474(01)00642-X}{\emph{Nucl. Phys.}
  {\bfseries A692} (2001) 583--645},
  [\href{https://arxiv.org/abs/hep-ph/0011241}{{\ttfamily hep-ph/0011241}}].

\bibitem{Ferreiro:2001qy}
E.~Ferreiro, E.~Iancu, A.~Leonidov and L.~McLerran, \emph{{Nonlinear gluon
  evolution in the color glass condensate. 2.}},
  \href{http://dx.doi.org/10.1016/S0375-9474(01)01329-X}{\emph{Nucl. Phys.}
  {\bfseries A703} (2002) 489--538},
  [\href{https://arxiv.org/abs/hep-ph/0109115}{{\ttfamily hep-ph/0109115}}].

\bibitem{Gelis:2010nm}
F.~Gelis, E.~Iancu, J.~Jalilian-Marian and R.~Venugopalan, \emph{{The Color
  Glass Condensate}},
  \href{http://dx.doi.org/10.1146/annurev.nucl.010909.083629}{\emph{Ann. Rev.
  Nucl. Part. Sci.} {\bfseries 60} (2010) 463--489},
  [\href{https://arxiv.org/abs/1002.0333}{{\ttfamily 1002.0333}}].

\bibitem{Fadin:1998py}
V.~S. Fadin and L.~Lipatov, \emph{{BFKL pomeron in the next-to-leading
  approximation}},
  \href{http://dx.doi.org/10.1016/S0370-2693(98)00473-0}{\emph{Phys. Lett. B}
  {\bfseries 429} (1998) 127--134},
  [\href{https://arxiv.org/abs/hep-ph/9802290}{{\ttfamily hep-ph/9802290}}].

\bibitem{Ciafaloni:1998gs}
M.~Ciafaloni and G.~Camici, \emph{{Energy scale(s) and next-to-leading BFKL
  equation}},
  \href{http://dx.doi.org/10.1016/S0370-2693(98)00551-6}{\emph{Phys. Lett. B}
  {\bfseries 430} (1998) 349--354},
  [\href{https://arxiv.org/abs/hep-ph/9803389}{{\ttfamily hep-ph/9803389}}].

\bibitem{Kotikov:2000pm}
A.~Kotikov and L.~Lipatov, \emph{{NLO corrections to the BFKL equation in QCD
  and in supersymmetric gauge theories}},
  \href{http://dx.doi.org/10.1016/S0550-3213(00)00329-1}{\emph{Nucl. Phys. B}
  {\bfseries 582} (2000) 19--43},
  [\href{https://arxiv.org/abs/hep-ph/0004008}{{\ttfamily hep-ph/0004008}}].

\bibitem{Balitsky:2008zza}
I.~Balitsky and G.~A. Chirilli, \emph{{Next-to-leading order evolution of color
  dipoles}}, \href{http://dx.doi.org/10.1103/PhysRevD.77.014019}{\emph{Phys.
  Rev. D} {\bfseries 77} (2008) 014019},
  [\href{https://arxiv.org/abs/0710.4330}{{\ttfamily 0710.4330}}].

\bibitem{Balitsky:2013fea}
I.~Balitsky and G.~A. Chirilli, \emph{{Rapidity evolution of Wilson lines at
  the next-to-leading order}},
  \href{http://dx.doi.org/10.1103/PhysRevD.88.111501}{\emph{Phys. Rev. D}
  {\bfseries 88} (2013) 111501},
  [\href{https://arxiv.org/abs/1309.7644}{{\ttfamily 1309.7644}}].

\bibitem{Kovner:2013ona}
A.~Kovner, M.~Lublinsky and Y.~Mulian, \emph{{Jalilian-Marian, Iancu, McLerran,
  Weigert, Leonidov, Kovner evolution at next to leading order}},
  \href{http://dx.doi.org/10.1103/PhysRevD.89.061704}{\emph{Phys. Rev. D}
  {\bfseries 89} (2014) 061704},
  [\href{https://arxiv.org/abs/1310.0378}{{\ttfamily 1310.0378}}].

\bibitem{Bartels:2002uz}
J.~Bartels, D.~Colferai, S.~Gieseke and A.~Kyrieleis, \emph{{NLO corrections to
  the photon impact factor: Combining real and virtual corrections}},
  \href{http://dx.doi.org/10.1103/PhysRevD.66.094017}{\emph{Phys. Rev. D}
  {\bfseries 66} (2002) 094017},
  [\href{https://arxiv.org/abs/hep-ph/0208130}{{\ttfamily hep-ph/0208130}}].

\bibitem{Balitsky:2012bs}
I.~Balitsky and G.~A. Chirilli, \emph{{Photon impact factor and
  $k_T$-factorization for DIS in the next-to-leading order}},
  \href{http://dx.doi.org/10.1103/PhysRevD.87.014013}{\emph{Phys. Rev. D}
  {\bfseries 87} (2013) 014013},
  [\href{https://arxiv.org/abs/1207.3844}{{\ttfamily 1207.3844}}].

\bibitem{Beuf:2016wdz}
G.~Beuf, \emph{{Dipole factorization for DIS at NLO: Loop correction to the
  $\gamma^*_{T,L}\to q\overline q$ light-front wave functions}},
  \href{http://dx.doi.org/10.1103/PhysRevD.94.054016}{\emph{Phys. Rev. D}
  {\bfseries 94} (2016) 054016},
  [\href{https://arxiv.org/abs/1606.00777}{{\ttfamily 1606.00777}}].

\bibitem{Boussarie:2016ogo}
R.~Boussarie, A.~Grabovsky, L.~Szymanowski and S.~Wallon, \emph{{On the one
  loop $ {\gamma}^{\left(\ast \right)}\to q\overline{q} $ impact factor and the
  exclusive diffractive cross sections for the production of two or three
  jets}}, \href{http://dx.doi.org/10.1007/JHEP11(2016)149}{\emph{JHEP}
  {\bfseries 11} (2016) 149},
  [\href{https://arxiv.org/abs/1606.00419}{{\ttfamily 1606.00419}}].

\bibitem{Chachamis:2013bwa}
G.~Chachamis, M.~Deak and G.~Rodrigo, \emph{{Heavy quark impact factor in
  kT-factorization}},
  \href{http://dx.doi.org/10.1007/JHEP12(2013)066}{\emph{JHEP} {\bfseries 12}
  (2013) 066}, [\href{https://arxiv.org/abs/1310.6611}{{\ttfamily 1310.6611}}].

\bibitem{Celiberto:2020tmb}
F.~G. Celiberto, D.~Y. Ivanov, M.~M. Mohammed and A.~Papa, \emph{{High-energy
  resummed distributions for the inclusive Higgs-plus-jet production at the
  LHC}},  \href{https://arxiv.org/abs/2008.00501}{{\ttfamily 2008.00501}}.

\bibitem{Chirilli:2011km}
G.~A. Chirilli, B.-W. Xiao and F.~Yuan, \emph{{One-loop Factorization for
  Inclusive Hadron Production in $pA$ Collisions in the Saturation Formalism}},
  \href{http://dx.doi.org/10.1103/PhysRevLett.108.122301}{\emph{Phys. Rev.
  Lett.} {\bfseries 108} (2012) 122301},
  [\href{https://arxiv.org/abs/1112.1061}{{\ttfamily 1112.1061}}].

\bibitem{Roy:2019hwr}
K.~Roy and R.~Venugopalan, \emph{{NLO impact factor for inclusive
  photon$+$dijet production in $e+A$ DIS at small $x$}},
  \href{http://dx.doi.org/10.1103/PhysRevD.101.034028}{\emph{Phys. Rev. D}
  {\bfseries 101} (2020) 034028},
  [\href{https://arxiv.org/abs/1911.04530}{{\ttfamily 1911.04530}}].

\bibitem{Hentschinski:2011tz}
M.~Hentschinski and A.~Sabio~Vera, \emph{{NLO jet vertex from Lipatov's QCD
  effective action}},
  \href{http://dx.doi.org/10.1103/PhysRevD.85.056006}{\emph{Phys. Rev. D}
  {\bfseries 85} (2012) 056006},
  [\href{https://arxiv.org/abs/1110.6741}{{\ttfamily 1110.6741}}].

\bibitem{Chachamis:2012gh}
G.~Chachamis, M.~Hentschinski, J.~Madrigal~Martinez and A.~Sabio~Vera,
  \emph{{Quark contribution to the gluon Regge trajectory at NLO from the high
  energy effective action}},
  \href{http://dx.doi.org/10.1016/j.nuclphysb.2012.03.015}{\emph{Nucl. Phys. B}
  {\bfseries 861} (2012) 133--144},
  [\href{https://arxiv.org/abs/1202.0649}{{\ttfamily 1202.0649}}].

\bibitem{Chachamis:2012cc}
G.~Chachamis, M.~Hentschinski, J.~D. Madrigal~Martínez and A.~Sabio~Vera,
  \emph{{Next-to-leading order corrections to the gluon-induced forward jet
  vertex from the high energy effective action}},
  \href{http://dx.doi.org/10.1103/PhysRevD.87.076009}{\emph{Phys. Rev. D}
  {\bfseries 87} (2013) 076009},
  [\href{https://arxiv.org/abs/1212.4992}{{\ttfamily 1212.4992}}].

\bibitem{Hentschinski:2014lma}
M.~Hentschinski, J.~Madrigal~Martínez, B.~Murdaca and A.~Sabio~Vera,
  \emph{{The next-to-leading order vertex for a forward jet plus a rapidity gap
  at high energies}},
  \href{http://dx.doi.org/10.1016/j.physletb.2014.06.022}{\emph{Phys. Lett. B}
  {\bfseries 735} (2014) 168--172},
  [\href{https://arxiv.org/abs/1404.2937}{{\ttfamily 1404.2937}}].

\bibitem{Nefedov:2017qzc}
M.~Nefedov and V.~Saleev, \emph{{On the one-loop calculations with Reggeized
  quarks}}, \href{http://dx.doi.org/10.1142/S0217732317502078}{\emph{Mod. Phys.
  Lett. A} {\bfseries 32} (2017) 1750207},
  [\href{https://arxiv.org/abs/1709.06246}{{\ttfamily 1709.06246}}].

\bibitem{Nefedov:2019mrg}
M.~A. Nefedov, \emph{{Computing one-loop corrections to effective vertices with
  two scales in the EFT for Multi-Regge processes in QCD}},
  \href{http://dx.doi.org/10.1016/j.nuclphysb.2019.114715}{\emph{Nucl. Phys. B}
  {\bfseries 946} (2019) 114715},
  [\href{https://arxiv.org/abs/1902.11030}{{\ttfamily 1902.11030}}].

\bibitem{Deak:2009xt}
M.~Deak, F.~Hautmann, H.~Jung and K.~Kutak, \emph{{Forward Jet Production at
  the Large Hadron Collider}},
  \href{http://dx.doi.org/10.1088/1126-6708/2009/09/121}{\emph{JHEP} {\bfseries
  09} (2009) 121}, [\href{https://arxiv.org/abs/0908.0538}{{\ttfamily
  0908.0538}}].

\bibitem{vanHameren:2013fla}
A.~van Hameren, P.~Kotko and K.~Kutak, \emph{{Three jet production and gluon
  saturation effects in p-p and p-Pb collisions within high-energy
  factorization}},
  \href{http://dx.doi.org/10.1103/PhysRevD.88.094001}{\emph{Phys. Rev. D}
  {\bfseries 88} (2013) 094001},
  [\href{https://arxiv.org/abs/1308.0452}{{\ttfamily 1308.0452}}].

\bibitem{vanHameren:2014ala}
A.~van Hameren, P.~Kotko, K.~Kutak and S.~Sapeta, \emph{{Small-$x$ dynamics in
  forward-central dijet decorrelations at the LHC}},
  \href{http://dx.doi.org/10.1016/j.physletb.2014.09.005}{\emph{Phys. Lett.}
  {\bfseries B737} (2014) 335--340},
  [\href{https://arxiv.org/abs/1404.6204}{{\ttfamily 1404.6204}}].

\bibitem{vanHameren:2014lna}
A.~van Hameren, P.~Kotko, K.~Kutak, C.~Marquet and S.~Sapeta, \emph{{Saturation
  effects in forward-forward dijet production in p$+$Pb collisions}},
  \href{http://dx.doi.org/10.1103/PhysRevD.89.094014}{\emph{Phys. Rev.}
  {\bfseries D89} (2014) 094014},
  [\href{https://arxiv.org/abs/1402.5065}{{\ttfamily 1402.5065}}].

\bibitem{vanHameren:2016ftb}
A.~van Hameren, P.~Kotko, K.~Kutak, C.~Marquet, E.~Petreska and S.~Sapeta,
  \emph{{Forward di-jet production in p+Pb collisions in the small-x improved
  TMD factorization framework}},
  \href{http://dx.doi.org/10.1007/JHEP12(2016)034}{\emph{JHEP} {\bfseries 12}
  (2016) 034}, [\href{https://arxiv.org/abs/1607.03121}{{\ttfamily
  1607.03121}}].

\bibitem{Bury:2016cue}
M.~Bury, M.~Deak, K.~Kutak and S.~Sapeta, \emph{{Single and double inclusive
  forward jet production at the LHC at $\sqrt{s}$ = 7 and 13 TeV}},
  \href{http://dx.doi.org/10.1016/j.physletb.2016.07.041}{\emph{Phys. Lett. B}
  {\bfseries 760} (2016) 594--601},
  [\href{https://arxiv.org/abs/1604.01305}{{\ttfamily 1604.01305}}].

\bibitem{Bury:2017xwd}
M.~Bury, H.~Van~Haevermaet, A.~Van~Hameren, P.~Van~Mechelen, K.~Kutak and
  M.~Serino, \emph{{Single inclusive jet production and the nuclear
  modification ratio at very forward rapidity in proton-lead collisions with
  $\sqrt{s_{NN}}$ = 5.02 TeV}},
  \href{http://dx.doi.org/10.1016/j.physletb.2018.03.007}{\emph{Phys. Lett. B}
  {\bfseries 780} (2018) 185--190},
  [\href{https://arxiv.org/abs/1712.08105}{{\ttfamily 1712.08105}}].

\bibitem{Kotko:2017oxg}
P.~Kotko, K.~Kutak, S.~Sapeta, A.~M. Stasto and M.~Strikman, \emph{{Estimating
  nonlinear effects in forward dijet production in ultra-peripheral heavy ion
  collisions at the LHC}},
  \href{http://dx.doi.org/10.1140/epjc/s10052-017-4906-6}{\emph{Eur. Phys. J.}
  {\bfseries C77} (2017) 353},
  [\href{https://arxiv.org/abs/1702.03063}{{\ttfamily 1702.03063}}].

\bibitem{Mantysaari:2019nnt}
H.~Mäntysaari and H.~Paukkunen, \emph{{Saturation and forward jets in
  proton-lead collisions at the LHC}},
  \href{http://dx.doi.org/10.1103/PhysRevD.100.114029}{\emph{Phys. Rev. D}
  {\bfseries 100} (2019) 114029},
  [\href{https://arxiv.org/abs/1910.13116}{{\ttfamily 1910.13116}}].

\bibitem{VanHaevermaet:2020rro}
H.~Van~Haevermaet, A.~Van~Hameren, P.~Kotko, K.~Kutak and P.~Van~Mechelen,
  \emph{{Trijets in kt-factorisation: matrix elements vs parton shower}},
  \href{https://arxiv.org/abs/2004.07551}{{\ttfamily 2004.07551}}.

\bibitem{Lipatov:1995pn}
L.~N. Lipatov, \emph{{Gauge invariant effective action for high-energy
  processes in QCD}},
  \href{http://dx.doi.org/10.1016/0550-3213(95)00390-E}{\emph{Nucl. Phys.}
  {\bfseries B452} (1995) 369--400},
  [\href{https://arxiv.org/abs/hep-ph/9502308}{{\ttfamily hep-ph/9502308}}].

\bibitem{Antonov:2004hh}
E.~N. Antonov, L.~N. Lipatov, E.~A. Kuraev and I.~O. Cherednikov,
  \emph{{Feynman rules for effective Regge action}},
  \href{http://dx.doi.org/10.1016/j.nuclphysb.2005.05.013,
  10.1016/j.nuclphysb.2005.013}{\emph{Nucl. Phys.} {\bfseries B721} (2005)
  111--135}, [\href{https://arxiv.org/abs/hep-ph/0411185}{{\ttfamily
  hep-ph/0411185}}].

\bibitem{vanHameren:2012uj}
A.~van Hameren, P.~Kotko and K.~Kutak, \emph{{Multi-gluon helicity amplitudes
  with one off-shell leg within high energy factorization}},
  \href{http://dx.doi.org/10.1007/JHEP12(2012)029}{\emph{JHEP} {\bfseries 12}
  (2012) 029}, [\href{https://arxiv.org/abs/1207.3332}{{\ttfamily 1207.3332}}].

\bibitem{vanHameren:2012if}
A.~van Hameren, P.~Kotko and K.~Kutak, \emph{{Helicity amplitudes for
  high-energy scattering}},
  \href{http://dx.doi.org/10.1007/JHEP01(2013)078}{\emph{JHEP} {\bfseries 01}
  (2013) 078}, [\href{https://arxiv.org/abs/1211.0961}{{\ttfamily 1211.0961}}].

\bibitem{Leonidov:1999nc}
A.~Leonidov and D.~Ostrovsky, \emph{{Angular and momentum asymmetry in particle
  production at high-energies}},
  \href{http://dx.doi.org/10.1103/PhysRevD.62.094009}{\emph{Phys. Rev. D}
  {\bfseries 62} (2000) 094009},
  [\href{https://arxiv.org/abs/hep-ph/9905496}{{\ttfamily hep-ph/9905496}}].

\bibitem{Kotko:2014aba}
P.~Kotko, \emph{{Wilson lines and gauge invariant off-shell amplitudes}},
  \href{http://dx.doi.org/10.1007/JHEP07(2014)128}{\emph{JHEP} {\bfseries 07}
  (2014) 128}, [\href{https://arxiv.org/abs/1403.4824}{{\ttfamily 1403.4824}}].

\bibitem{vanHameren:2016kkz}
A.~van Hameren, \emph{{KaTie : For parton-level event generation with
  $k_T$-dependent initial states}},
  \href{http://dx.doi.org/10.1016/j.cpc.2017.11.005}{\emph{Comput. Phys.
  Commun.} {\bfseries 224} (2018) 371--380},
  [\href{https://arxiv.org/abs/1611.00680}{{\ttfamily 1611.00680}}].

\bibitem{vanHameren:2017hxx}
A.~van Hameren, \emph{{Calculating off-shell one-loop amplitudes for
  $k_T$-dependent factorization: a proof of concept}},
  \href{https://arxiv.org/abs/1710.07609}{{\ttfamily 1710.07609}}.

\bibitem{Dumitru:2005gt}
A.~Dumitru, A.~Hayashigaki and J.~Jalilian-Marian, \emph{{The Color glass
  condensate and hadron production in the forward region}},
  \href{http://dx.doi.org/10.1016/j.nuclphysa.2005.11.014}{\emph{Nucl. Phys.}
  {\bfseries A765} (2006) 464--482},
  [\href{https://arxiv.org/abs/hep-ph/0506308}{{\ttfamily hep-ph/0506308}}].

\bibitem{Mangano:1990by}
M.~L. Mangano and S.~J. Parke, \emph{{Multiparton amplitudes in gauge
  theories}}, \href{http://dx.doi.org/10.1016/0370-1573(91)90091-Y}{\emph{Phys.
  Rept.} {\bfseries 200} (1991) 301--367},
  [\href{https://arxiv.org/abs/hep-th/0509223}{{\ttfamily hep-th/0509223}}].

\bibitem{bern1993new}
Z.~Bern, L.~Dixon and D.~A. Kosower, \emph{New qcd results from string theory},
   tech. rep., 1993.

\bibitem{bern1994one}
Z.~Bern, G.~Chalmers, L.~Dixon and D.~A. Kosower, \emph{One-loop n-gluon
  amplitudes with maximal helicity violation via collinear limits},
  {\emph{Physical review letters} {\bfseries 72} (1994) 2134}.

\bibitem{mahlon1994multigluon}
G.~Mahlon, \emph{Multigluon helicity amplitudes involving a quark loop},
  {\emph{Physical Review D} {\bfseries 49} (1994) 4438}.

\bibitem{Bern_2011}
Z.~Bern and Y.~tin Huang, \emph{Basics of generalized unitarity},
  \href{http://dx.doi.org/10.1088/1751-8113/44/45/454003}{\emph{Journal of
  Physics A: Mathematical and Theoretical} {\bfseries 44} (2011) 454003}.

\bibitem{Arkani-Hamed_book_2016}
N.~Arkani-Hamed, J.~Bourjaily, F.~Cachazo, A.~Goncharov, A.~Postnikov and
  J.~Trnka, \emph{Grassmannian Geometry of Scattering Amplitudes}.
\newblock Cambridge University Press, 2016,
  \href{http://dx.doi.org/10.1017/CBO9781316091548}{10.1017/CBO9781316091548}.

\bibitem{bern1993one}
Z.~Bern, L.~Dixon and D.~A. Kosower, \emph{One-loop corrections to five-gluon
  amplitudes}, {\emph{Physical Review Letters} {\bfseries 70} (1993) 2677}.

\bibitem{bern1995one}
Z.~Bern, L.~Dixon and D.~A. Kosower, \emph{One-loop corrections to two-quark
  three-gluon amplitudes}, {\emph{Nuclear Physics B} {\bfseries 437} (1995)
  259--304}.

\bibitem{nefedov2019one}
M.~Nefedov, \emph{One-loop corrections to multiscale effective vertices in the
  eft for multi-regge processes in qcd},
  \href{https://arxiv.org/abs/1905.01105}{{\ttfamily 1905.01105}}.

\bibitem{bern2005last}
Z.~Bern, L.~J. Dixon and D.~A. Kosower, \emph{Last of the finite loop
  amplitudes in qcd}, {\emph{Physical Review D} {\bfseries 72} (2005) 125003}.

\bibitem{vanHameren:2014iua}
A.~van Hameren, \emph{{BCFW recursion for off-shell gluons}},
  \href{http://dx.doi.org/10.1007/JHEP07(2014)138}{\emph{JHEP} {\bfseries 07}
  (2014) 138}, [\href{https://arxiv.org/abs/1404.7818}{{\ttfamily 1404.7818}}].

\bibitem{Kotko:2015ura}
P.~Kotko, K.~Kutak, C.~Marquet, E.~Petreska, S.~Sapeta and A.~van Hameren,
  \emph{{Improved TMD factorization for forward dijet production in
  dilute-dense hadronic collisions}},
  \href{http://dx.doi.org/10.1007/JHEP09(2015)106}{\emph{JHEP} {\bfseries 09}
  (2015) 106}, [\href{https://arxiv.org/abs/1503.03421}{{\ttfamily
  1503.03421}}].

\bibitem{Altinoluk:2019fui}
T.~Altinoluk, R.~Boussarie and P.~Kotko, \emph{{Interplay of the CGC and TMD
  frameworks to all orders in kinematic twist}},
  \href{http://dx.doi.org/10.1007/JHEP05(2019)156}{\emph{JHEP} {\bfseries 05}
  (2019) 156}, [\href{https://arxiv.org/abs/1901.01175}{{\ttfamily
  1901.01175}}].

\end{thebibliography}\endgroup
\end{document}